\documentclass[journal=pasa, manuscript=research-paper, year=2025, volume=37]{cup-journal}
\usepackage{graphicx}
\usepackage{amsmath}
\usepackage[nopatch]{microtype}
\usepackage{booktabs}
\usepackage{hyperref}
\usepackage{url}
\usepackage{graphics}
\usepackage{xurl}
\usepackage[dvipsnames]{xcolor}
\usepackage{balance}
\usepackage{soul}
\usepackage[font=footnotesize,labelformat=simple]{subcaption}

\usepackage[T1]{fontenc}
\usepackage{xspace}
\usepackage{tikz}
\usepackage{amsmath,amssymb}
\usepackage{bm}
\newcommand{\hcon}{\ensuremath{\mathrm{km~s^{-1}~Mpc^{-1}}}\xspace}

\title{Dark sirens and the impact of redshift precision}

\author{Madeline L. Cross-Parkin}
\affiliation{School of Mathematics and Physics, University of Queensland, Brisbane, QLD 4072, Australia}
\alsoaffiliation{OzGrav: The ARC Centre of Excellence for Gravitational Wave Discovery, Australia}
\email[M.L. Cross-Parkin]{m.crossparkin@uq.edu.au}

\author{Cullan Howlett}
\affiliation{School of Mathematics and Physics, University of Queensland, Brisbane, QLD 4072, Australia}
\alsoaffiliation{OzGrav: The ARC Centre of Excellence for Gravitational Wave Discovery, Australia}
\author{Tamara M. Davis}
\affiliation{School of Mathematics and Physics, University of Queensland, Brisbane, QLD 4072, Australia}
\alsoaffiliation{OzGrav: The ARC Centre of Excellence for Gravitational Wave Discovery, Australia}
\author{Nandita Khetan}
\affiliation{School of Mathematics and Physics, University of Queensland, Brisbane, QLD 4072, Australia}
\alsoaffiliation{OzGrav: The ARC Centre of Excellence for Gravitational Wave Discovery, Australia}


\begin{document}
\begin{abstract}
    With the growing number of gravitational wave detections, achieving a competitive measurement of $H_0$ with dark sirens is becoming increasingly feasible. The expansion of the LIGO-Virgo-KAGRA Collaboration into a four detector network will reduce both the localisation area and the luminosity distance uncertainty associated with each gravitational wave event. It is therefore essential to identify and mitigate other major sources of error that could increase the uncertainty in $H_0$. In this work, we explore three scenarios relevant to the dark siren method in future observing runs. First, we demonstrate that there is a precision gain offered by a catalogue of spectroscopic-like redshifts compared to photometric-like redshifts, with the greatest improvements observed in smaller localisation areas. Second, we show that redshift outliers (as occur in realistic photometric redshift catalogues), do not introduce bias into the measurement of $H_0$. Finally, we find that uniformly sub-sampling spectroscopic-like redshift catalogues increases the uncertainty in $H_0$ as the completeness fraction is decreased; at a completeness of 50\% the benefit of spectroscopic redshift precision is outweighed by the degradation from incompleteness. In all three scenarios, we obtain unbiased estimates of $H_0$. We conclude that a competitive measurement of $H_0$ using the dark siren method will require a hybrid catalogue of both photometric and spectroscopic redshifts, at least until highly complete spectroscopic catalogues become available. This, however, will come at the cost of a more complex selection function.
\end{abstract}

\section{Introduction}
\label{sec:intro}
Gravitational waves from compact binary mergers provide a powerful method for measuring the expansion rate of the Universe, quantified by the Hubble constant ($H_0$). This so-called standard siren analysis \citep{Schutz} has been performed on multiple occasions following the first detection of a gravitational wave in 2015 \citep{PhysRevLett.116.061102} by the Laser Interferometer Gravitational Wave Observatory (LIGO; \citealt{2015CQGra..32g4001L}). Since this inaugural detection, the network has expanded to include two additional detectors: Virgo \citep{2015CQGra..32b4001A} and KAGRA \citep{2019}. This expanded network enhances the triangulation of gravitational wave signals by utilising arrival time delays across multiple detectors, significantly reducing the localisation area for each event \citep{2009NJPh...11l3006F}.

Gravitational waves are powerful cosmological probes because they serve as absolute distance indicators, eliminating the need for a distance-ladder calibration to infer cosmological distances. However, constraining cosmological parameters requires an additional independent measurement of the source redshift to establish a connection with the luminosity distance. More specifically, determining $H_0$ involves using the redshift to calculate the recession velocity, \footnote{The equation for recession velocity is $v(z)=c\int_0^z \frac{dz}{E(z)}$, where $E(z)$ is the Hubble parameter divided by its value at the present day $H(z)/H_0$; at low redshifts ($z\ll 0.1$) this can be approximated by a cosmographic expansion, see Equation~14 of \citet{Davis2014}.} $v(z)$, alongside a distance measurement, ideally the proper distance, $D$. The proper distance is the distance used in the Hubble-Lemaître Law, $v=H_0D$ \citep{1931MNRAS..91..490L, 1929PNAS...15..168H}. 

The Hubble constant has been a parameter of interest for almost a century, but despite the plethora of different approaches to measuring $H_0$ \citep{1993ApJ...417..553L,Freedman_2020,Planck,Freedman_2021,lee2024chicagocarnegiehubbleprogramjwst,Riess,osti_2346175}, there remains a significant discrepancy between high- and low-redshift measurements. Most notably, measurements obtained with Cepheid variable calibrated Type Ia supernovae \citep{Riess} and the Cosmic Microwave Background \citep{Planck} have both reached precision better than $2\%$, however the two measurements are in tension between $4\sigma$ and $5.8\sigma$ \citep{Verde_2019}. This Hubble tension necessitates new independent measurements of $H_0$, which could clarify the source of this tension --- whether it originates from unknown systematic errors or beyond $\Lambda$CDM physics \citep{2017NatAs...1E.169F}.

Standard siren analysis refers to the process of using gravitational waves to measure $H_0$. It was applied for the first time to the binary neutron star merger GW170817 \citep{PhysRevLett.119.161101, 2017Natur.551...85A} in which the corresponding electromagnetic counterpart \citep{2017ApJ...848L..12A} allowed for the identification of the host galaxy, leading to a $\sim 20\%$ precision measurement of $H_0$. However, the dominant source of uncertainty in this measurement arose from the degeneracy between the source distance and its orbital inclination, leading to significant uncertainties in the inferred distance. \citet{hotokezaka2018hubbleconstantmeasurementsuperluminal} mitigated this by analysing the superluminal motion of the jet associated with GW170817 using Very Long Baseline Interferometry (VLBI). By constraining the viewing angle more precisely, they refined the $H_0$ measurement to a precision of approximately 7\%. 

Measurements of $H_0$ that utilise the electromagnetic counterpart alongside the gravitational wave signal are referred to as bright sirens. Without a counterpart, it is still possible to measure $H_0$ if there exists a catalogue of galaxies within the localisation region of the gravitational wave. \color{Black}This dark siren (or statistical siren) method combines the gravitational wave signal with the redshift distribution of galaxies within the localisation region to weigh the likelihood of each galaxy being the host of the merger \citep{Holz_2005,PhysRevD.86.043011, Chen:2017rfc,LIGOScientific:2018gmd, Soares_Santos_2019, Gray:2019ksv, Finke:2021aom, LIGOScientific:2021aug, 2023AJ....166...22G}. \color{Black}This information is then used to constrain $H_0$ by relating the distribution of galaxy redshifts to the luminosity distance of the gravitational wave and averaging over many such sources. 

\color{Black}In order to obtain realistic and unbiased constraints, any dark siren analyses must account for the underlying binary black hole mass distribution and its possible evolution with redshift \citep{gray2023jointcosmologicalgravitationalwavepopulation, PhysRevD.108.042002, Borghi_2024}. In this work, we adopt the simplifying assumption of not marginalising over the binary black hole mass distribution (as also done in \citealt{2023AJ....166...22G}), as our goal is to isolate systematics associated with the galaxy catalogue rather than to model the mass function. A more correct analysis that jointly incorporates the mass distribution and its redshift evolution is left for future work.

%In the galaxy catalogue approach --- where host galaxies are statistically associated with gravitational wave events but detector-frame masses are not used --- a realistic analysis would require one to draw events from an underlying mass distribution in order to capture mass-dependent selection effects. Spectral siren analyses --- which use detector-frame masses to infer source redshifts, are even more sensitive to the choice of the underlying binary black hole mass distribution. In such cases, accurately modelling the mass distribution is essential to avoid introducing biases into the inferred cosmological parameters. We note that this work does not include marginalisation over the binary black hole mass distribution. Our goal is to isolate galaxy catalogue systematics rather than model the mass function, and so we uniformly sample gravitational wave hosts. A more comprehensive treatment that incorporates the binary black hole mass distribution and its possible redshift evolution is left for future work.

\color{Black}While bright sirens offer tighter constraints for any single event, \citet{Palmese_2023} shows that 8 well-localised dark sirens combined with an appropriate galaxy catalogue is able to provide a $H_0$ constraint that is competitive with a single bright siren measurement. Their sample contains the top 20\% (based on localisation area) gravitational wave events to date, and they make use of a combination of both spectroscopy (from DESI; \citealt{2016arXiv161100036D}) and photometry (from DES; \citealt{2005astro.ph.10346T}). While they do not marginalise over the mass distribution, instead adopting a fixed one, they find that their $H_0$ results are largely insensitive to variations in the assumed mass distribution for their selected sample of events. \color{Black}Their measurement for the 8 dark sirens yields $H_0=79.8^{+19.1}_{-12.8}$ \hcon, achieving a precision of 18\% compared to the 20\% precision from the bright siren GW170817 when no inclination angle information is included. Given that we have detected a single bright siren and $\mathcal{O}(100)$ dark sirens, improving the modelling associated with dark siren analysis is likely to overall yield a tighter constraint on $H_0$.

Due to the nature of dark siren analysis, the redshift uncertainties of galaxies within the localisation area directly influence the precision of the measured Hubble constant. This analysis aims to identify the regimes in which the greater precision of spectroscopic redshifts is most beneficial for measuring $H_0$, compared to photometric redshifts. There will be a trade-off in that photometric redshift catalogues are more complete than spectroscopic catalogues, so we have to investigate both how redshift precision and completeness affect our results. In detail, we ask:
\begin{enumerate}[label={\roman*}.]
    \item \label{itme:q1}Does the increased precision of spectroscopic redshifts improve the constraints on $H_0$ (assuming an equal sized sample), and does this improvement depend on the size of the localisation region? 
    \item  How will biases from photometric-redshift outliers propagate into $H_0$ measurements and will these biases impact our understanding of the Hubble tension? 
    \item At what level of incompleteness does the degradation of accuracy due to incompleteness outweigh the higher-precision of spectroscopic redshifts?
\end{enumerate}
To answer these questions we split our analysis into three parts:
\begin{enumerate}
\item \label{item:1} Firstly, in our basic analysis we assume Gaussian uncertainties on both spectroscopic and photometric redshifts. 
\item Secondly, we consider more realistic photometric redshift distributions that include catastrophic redshift failures (outliers), leading to non-Gaussian uncertainties.
\item Thirdly, we use a similar methodology to Point~\ref{item:1} but remove a percentage of the galaxies after assigning gravitational wave hosts to simulate what an incomplete redshift catalogue could look like.
\end{enumerate}

\citet{turski2023impactmodellinggalaxyredshift} is complimentary to this work, as it explores the effect of galaxy redshift uncertainties on a measurement of $H_0$ with current gravitational wave detection capabilities. They find that the potential bias introduced by using photometric redshift catalogues is significantly smaller than the current statistical and systematic uncertainties associated with existing detections. Our work considers the prospects for future dark siren analyses, where both the number of detections is larger, and fractional luminosity distance uncertainty for each detection is smaller. Additionally, while \citet{turski2023impactmodellinggalaxyredshift} primarily explores redshift uncertainty models, our study highlights the combined impact of gravitational wave localisation area and redshift uncertainty (spectroscopic versus photometric) on the precision of future $H_0$ measurements.

\color{Black}In addition, \citet{Borghi_2024} address a similar question to \ref{itme:q1}, investigating the impact of no-outlier photometric and spectroscopic redshift catalogues on $H_0$ measurements. They find that the use of a photometric redshift catalogue degrades the precision of $H_0$ constraints by a factor of $\sim3$ compared to a spectroscopic catalogue. However, in contrast to our methodology, they combine a spectral siren analysis with the galaxy catalogue approach; our analysis uses only the latter.

\color{Black}This paper is structured as follows. In Section~\ref{sec:motivation} we present the motivation behind our three aims. We describe the data used in Section~\ref{sec:galcat} and the Bayesian framework and extensions to this model in Section~\ref{sec:statframework}. Our results and discussion from our three aims follows in Sections~\ref{sec:resultsgaussian}, \ref{sec:resultsrealistic} and \ref{sec:completeness}, and we conclude in Section~\ref{sec:conclusions}.

\section{The problem and motivations}
\label{sec:motivation}
\begin{figure}[t]
    \centering
    \includegraphics[width=0.98\linewidth]{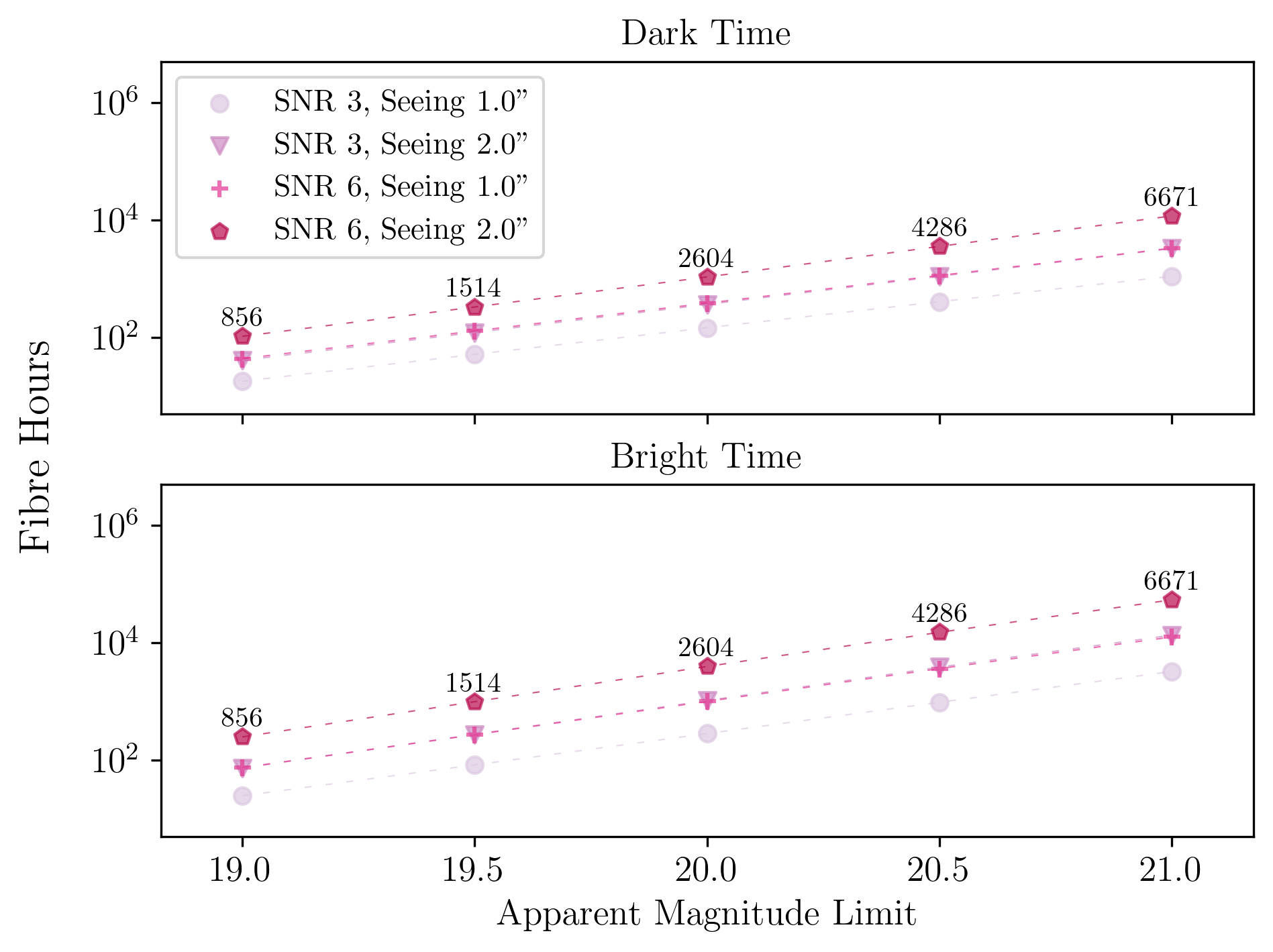}
    \caption{The fibre hours required on a telescope like the AAT to spectroscopically observe all galaxies individually within a one-square-degree localisation area are shown for five different apparent magnitude limits. These fibre hours represent only the exposure time and therefore exclude configuration time. The top panel presents exposure times for four combinations of signal-to-noise ratio and seeing conditions under dark time, while the bottom panel displays the same for bright time conditions. The numbers above each scatter point indicate the total number of galaxies within the localisation area that fall below the corresponding apparent magnitude limit. Note: the SNR 3, Seeing 2.0" are hidden behind the SNR 6, Seeing 1.0" points.}
    \label{fig:ExpTime}
\end{figure}

This section provides a benchmark of the observational time required to spectroscopically survey all galaxies within a gravitational wave localisation area and below an apparent magnitude limit. To estimate the number of galaxies in a 1 deg$^2$ localisation area, we use the $r$-band luminosity function as measured by \citet{Loveday_2011}. Figure~\ref{fig:ExpTime} illustrates the estimated number of galaxies for a range of apparent magnitude limits, along with the corresponding exposure time required to spectroscopically observe each galaxy individually with a 4~m telescope like the Anglo-Australian Telescope (AAT; \citealt{AAT}). 
\citet{Childress_2017} finds that the redshift success rate increases significantly at a signal-to-noise of 2 to 3 per 1-Å bin; we use this as the baseline SNR. Note that Figure~\ref{fig:ExpTime} shows the total exposure time to observe each galaxy individually and thus does not take into account multiplexing, where multiple objects can be observed simultaneously during a single exposure. Additionally, these values represent only the exposure times and do not account for configuration times or field-of-view considerations.

Using the AAT as an example, with a fibre density of 101 fibres per square degree, a 2-degree diameter field of view, and a configuration time of 40 minutes, observing all galaxies within a single square-degree localisation area to an apparent magnitude of $19.0$ under ideal conditions (dark time, 1.0" seeing, and SNR/Å of 3) requires $\sim6$ hours. For an apparent magnitude limit of $21.0$, corresponding to 6671 galaxies, the total observational time increases to $\sim45$ hours. In both such cases, the configuration time is the limiting factor in the number of redshifts that can be obtained.\footnote{For galaxies with $m\leq21.0$, the exposure time under ideal conditions is shorter than the configuration time (keep in mind the AAT can configure and expose simultaneously), meaning the total observation time is determined by the product of the configuration time and the number of pointings.} 

By contrast, modern instruments on 4-meter-class telescopes like 4MOST \citep{2012SPIE.8446E..0TD} and DESI \citep{DESI_Collaboration_2022} are much more efficient due to their advanced multi-fibre systems and shorter configuration times. DESI, for example, has a fibre density of 621 fibres per square degree, a 3.2-diameter field of view and a configuration time of 2 minutes \citep{desicollaboration2016desiexperimentiiinstrument}. Unlike the AAT, where configuration and exposure are performed simultaneously, DESI configures and exposes sequentially. Nonetheless, its short configuration time renders this process highly effective, but still takes $\sim4.7$ hours to observe all galaxies for $m\leq 21.0$ in a single square degree.\footnote{These estimates likely represent an upper limit, as they do not account for the improved throughput and spectral performance of modern instruments compared to 2dF.} \color{Black}Thus, although DESI represents a major improvement over earlier-generation facilities and is no longer limited by configuration time, the total observing time remains substantial even for very small localisation areas. For more realistic localisation areas and deeper magnitude limits, the required observing time would increase significantly. Completing such observations would require many days of telescope time, making them both expensive and difficult to justify for time allocation. Moreover, DESI and 4MOST will, for the foreseeable future, operate exclusively as survey instruments, without the possibility of securing dedicated observing time for targeted follow-up.

\color{Black}This limitation in observational efficiency becomes particularly relevant in the context of host galaxy completeness for gravitational wave events. \citet{Rauf_2023} examines the binary black hole merger rate completeness as a function of distance for various $r$-band apparent magnitudes. For a maximum distance of 3~Gpc --- encompassing all current detections with 50\% localisation areas below 10 deg$^2$ --- only 25-35\% of binary black hole merger host galaxies would be observable with an apparent magnitude limit of 21.0.\footnote{\color{Black}This exact number may vary depending on the assumed model of binary black hole evolution.} \color{Black}Naturally, this completeness would improve for closer gravitational wave events or a higher apparent magnitude cutoff.

Even for small localisation areas and under ideal conditions, the total observation time grows rapidly as the apparent magnitude limit increases. This motivates an evaluation into whether the improved precision afforded by spectroscopic redshifts justifies the additional observing time, particularly given the potential limitations imposed by biases from photometric outliers and the effects of catalogue incompleteness on the accuracy of $H_0$ measurements.

\section{Galaxy catalogues}
\label{sec:galcat}
Galaxy catalogues are essential for dark siren analyses; they provide the redshift information necessary to complement the gravitational wave's luminosity distance, enabling the identification of potential host galaxies and a measurement of $H_0$.

\subsection{MICEcat}
To generate mock gravitational wave events and observed redshifts, we use galaxies taken from the MICEcat mock galaxy catalogue \citep{Fosalba_2015}, a light-cone simulation covering one-eighth of the sky out to a redshift of $z=1.4$. The fiducial cosmological model used in MICEcat is a flat $\Lambda$CDM model, with cosmological parameters $H_0=70$ \hcon, $\Omega_m=0.25$, $\Omega_b=0.044$, $\Omega_\Lambda=0.75$. A subsample of the MICEcat data is used for this analysis to reduce the number of galaxies to $\sim 270$ galaxies per square degree. This choice will not introduce any bias as all host galaxies are being selected from this subsample. Moreover, tests confirm that the subsample produces results consistent with the full dataset while significantly improving computational efficiency.

\subsection{redMaGiC}
\label{sec:redMaGiC}
To assess the accuracy of photometric redshifts, we use data collected by Dark Energy Survey, specifically the redMaGiC sample \citep{Rozo_2016}. The redMaGiC algorithm was designed to minimise photometric redshift uncertainties in a sample of Luminous Red Galaxies. It was chosen for this analysis as a benchmark for the capabilities of current state-of-the-art and anticipated future surveys. The redMaGiC catalogue includes galaxies with both photometric and spectroscopic redshifts, primarily concentrated in the range $z\in[0.15, 0.7]$. To ensure consistent comparisons, we filter MICEcat data to align with this interval.

\begin{figure}[t]
    \centering
    \includegraphics[width=\linewidth]{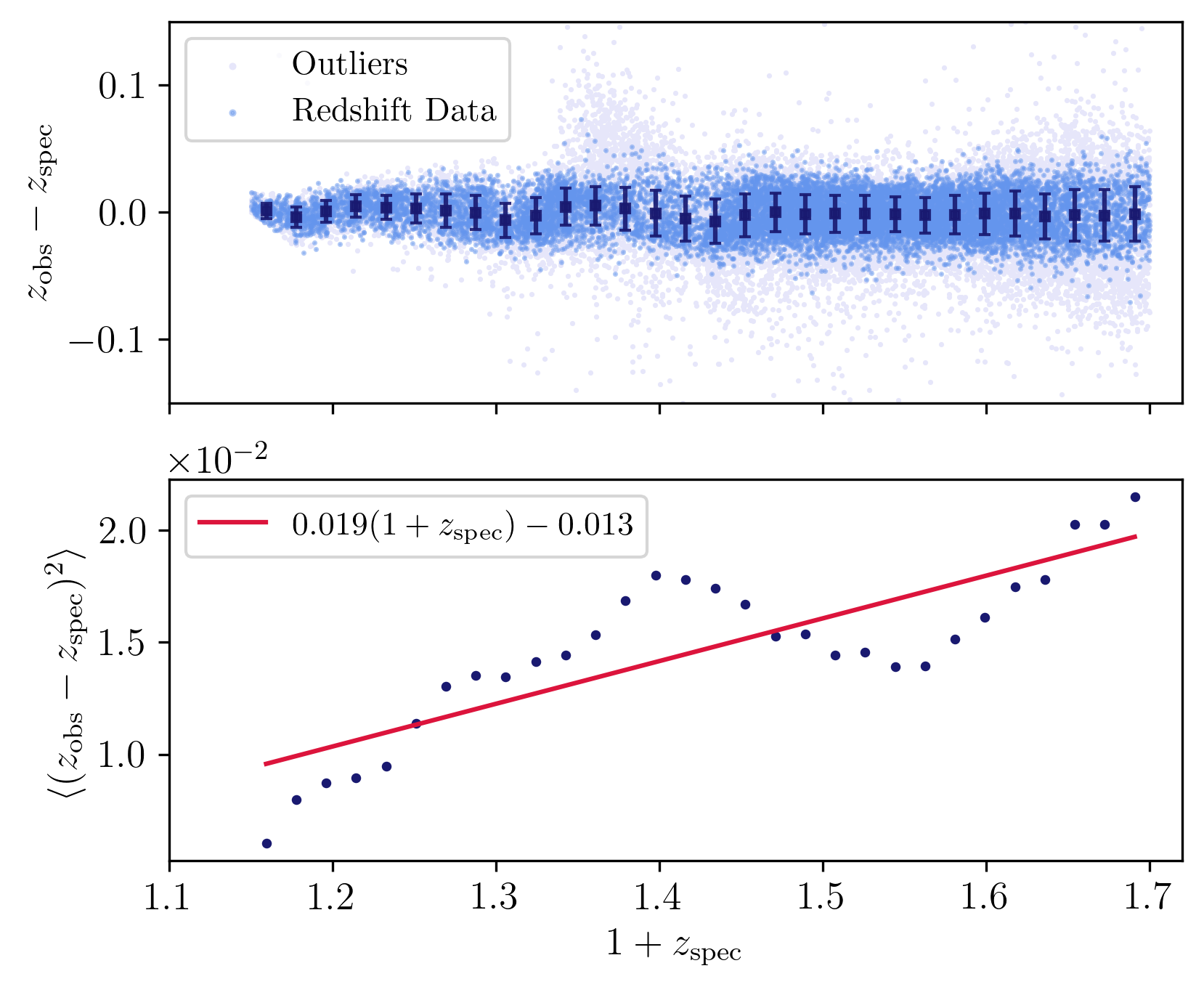}
    \caption{The top panel shows the redMaGiC redshift sample of photometric redshifts minus the corresponding spectroscopic redshifts. Outliers are identified by calculating the $p$-value of the $\chi^2$ statistic between the photometric and spectroscopic redshift for each galaxy and removing the galaxy using the $p$-value as a sampling probability. The remaining redshifts are binned, each bin having a mean and standard deviation. The bottom panel shows the standard deviation of each bin, with a best fit line that relates the spectroscopic redshift to the standard deviation.}
    \label{fig:DESResidual}
\end{figure}

After excluding outliers using sigma-clipping,\footnote{Outliers are excluded only for the purpose of establishing the relationship between redMaGiC spectroscopic and photometric redshifts. Once this relationship is determined, the outliers are reintroduced into the catalogue.} we establish an approximate relationship between the redMaGiC spectroscopic and photometric redshifts by binning the redshifts and computing the mean and standard deviation within each bin. \color{Black}The line of best fit through these standard deviations characterises the relationship, which can be seen in Figure~\ref{fig:DESResidual}. The oscillations present in Figure~\ref{fig:DESResidual} about the best fit arise from structure present in the photometric redshift catalogues, likely due to the methods used to estimate redshifts from training data and input photometry. In particular, strong spectral features shift in and out of photometric filter bands with redshift, leading to discontinuities or artificial clustering in the estimated redshift distribution.

\color{Black}We find that the standard deviation follows roughly the relationship
 \begin{equation}
 \label{eq:DESstdev}
 	\text{standard deviation} = \sigma_z(1+z_{\mathrm{spec}})+A.
 \end{equation}
 
This relationship is applied in our first and third aims to generate observed redshifts from the true redshifts in MICEcat by setting $A=0$ and varying $\sigma_z$. For the second aim, we use the best fit parameters to the redMaGiC data, as shown in Figure~\ref{fig:DESResidual}, with $A=-0.013$ and $\sigma_z=0.019$. We then compare the results from this catalogue to those obtained when directly using the redMaGiC redshifts as the observed redshifts.

\section{Statistical framework}
\label{sec:statframework}
Our analysis is based on the methodology and code provided in \citet{2023AJ....166...22G} which we have modified for our purposes.\footnote{\url{https://github.com/simone-mastrogiovanni/hitchhiker_guide_dark_sirens}}

Assuming we have a number of gravitational wave events with observed data $\{\hat{x}\}$, by adopting a simplified approach where the observed data includes only the observed gravitational wave luminosity distances, $\{\hat{d}_L$\}, the posterior on $H_0$ can be expressed using Bayes' theorem as
\begin{equation}
    p(H_0 \mid\{\hat{d}_L\}) \propto \mathcal{L}(\{\hat{d}_L\} \mid H_0) \ p(H_0),
\end{equation}
\noindent where $p(H_0)$ is the prior on $H_0$, taken to be uniform over the range $H_0\in[40,100]$ \hcon and $\mathcal{L}(\{\hat{d}_L\} \mid H_0)$ is the likelihood of observing the data $\{\hat{d}_L\}$ given a value for $H_0$. In this analysis we distinguish observed from true parameters with `hats' ($\hat{\phantom{x}}$). To reduce ambiguity, a summary of all parameters referred to in the following framework is shown in Table~\ref{tab:params}.
\begin{table}[t!]
\centering
\renewcommand{\arraystretch}{1.5}
\resizebox{\columnwidth}{!}{%
\begin{tabular}{cc}
\textbf{Parameter} & \textbf{Description}\\
\midrule
$p(H_0 \mid\{\hat{d}_L\})$  &  $H_0$ posterior given measured luminosity distances \\
$\mathcal{L}(\{\hat{d}_L\} \mid H_0)$ & Likelihood of $H_0$ given measured luminosity distances\\
$p(H_0)$ & Prior on $H_0$  \\
$\mathcal{L}_{\mathrm{GW}}(\hat{d}_L^{i} \mid d_L(z, H_0))$ & Gaussian gravitational wave likelihood\\
$p_{\mathrm{CBC}}(z)$ & Probability that a CBC will occur at redshift $z$ \\
$p_{\mathrm{rate}}(z)$ &Probability of a galaxy at redshift $z$ hosting a merger \\
$p_{\mathrm{cat}}(z)$ & Probability a galaxy exists at redshift $z$\\
$p_{\mathrm{det}}^{\mathrm{GW}}(z,H_0)$ & Probability of detecting a gravitational wave\\
$p_{\mathrm{red}}(z \mid \hat{z}_g^i)$ & True redshift posterior given measured redshifts\\
$\mathcal{L}_{\mathrm{red}}(\hat{z}_g^i \mid z)$ & Probability of observing a redshift given a true redshift\\
$p_{\mathrm{bg}}(z)$ & Prior on the redshift distribution\\
$\hat{d}_L(z,H_0)$ & Observed luminosity distance of the GW\\
$\sigma_{d_L}$ & Uncertainty in the luminosity distance\\
$\hat{d}_L^{\mathrm{thr}}$ & Threshold at which a gravitational wave can be observed \\
$z_\mathrm{lower}$ and $z_\mathrm{upper}$ & True redshift cutoffs\\
$\hat{z}_g$, $\hat{d}_L$ & Observed redshift/luminosity distance\\
$N_{\mathrm{gal}}$ & Number of galaxies in the localisation area \\
\bottomrule
\end{tabular}}
\caption{Summary of the parameters referred to in this framework. Their functional forms are defined throughout Section~\ref{sec:statframework}.}
\label{tab:params}
\end{table}
The likelihood of each individual luminosity distance measurement can be written as:\color{Black}
\begin{align}
        \mathcal{L}(\hat{d}_L^i \mid H_0)=\frac{\iint d\Omega\ dz \ \mathcal{L}_{\mathrm{GW}}(\hat{d}_L^{i} \mid d_L(z, H_0), \Omega) \ p_{\mathrm{CBC}}(z,\Omega)}{\iint d\Omega\ dz \ p_{\mathrm{det}}^{\mathrm{GW}}(z,H_0, \Omega) \ p_{\mathrm{CBC}}(z, \Omega)},
\end{align}
\noindent where $\mathcal{L}_{\mathrm{GW}}$ represents the likelihood of observing a luminosity distance $\hat{d}_L^i$, given a galaxy with a true luminosity distance $d_L(z, H_0)$. Here, $\Omega$ denotes the solid angle on the sky. Since we assume a uniform probability distribution within each localisation area --- every galaxy within the cone is assigned equal weight --- the angular dependence factors out, and the expression simplifies to an integral over redshift only. Future work will explore the impact of using more realistic probability maps, such as those with Gaussian profiles. 

\color{Black}Assuming a Gaussian form for the gravitational wave likelihood, this can be expressed as:
\begin{equation}
    \label{eq:LGW}
    \mathcal{L}_{\mathrm{GW}}\left(\hat{d}_L^i \mid d_L\left(z, H_0\right)\right) = \mathcal{N}\left(d_L\left(z, H_0\right)\mid\sigma_{d_L}\right),
\end{equation}
\noindent where $\sigma_{d_L}$ is the gravitational wave luminosity distance uncertainty. 

\begin{figure}[t!]
    \centering
    \includegraphics[width=\linewidth]{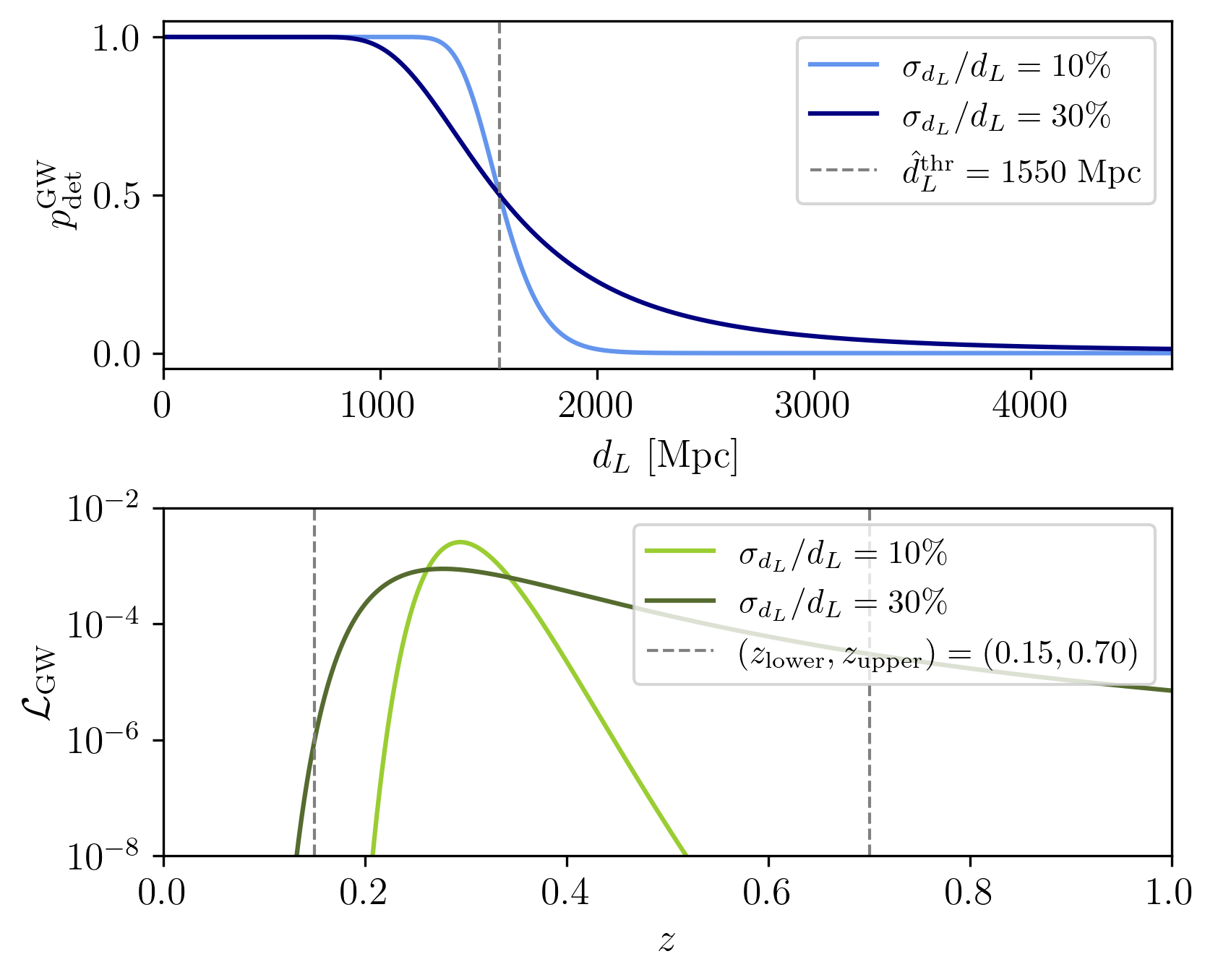}
    \caption{The top panel shows the probability of detecting a gravitational wave at LIGO (given 2021 LVK capabilities) as a function of luminosity distance for two fractional luminosity distance uncertainties. In practice, the detection probability depends on the specific sensitivity of the gravitational wave detector. \color{Black}Unless otherwise stated, we simulate only gravitational wave events with a fractional luminosity distance uncertainty of $\sigma_{d_L}/d_L=10\%$ in this analysis. \color{Black}The dashed line at $\hat{d}_L^{\mathrm{thr}}=1550$ Mpc represents the threshold assumed in our analysis, beyond which galaxies with higher observed luminosity distances are considered undetectable (essentially acting as a signal-to-noise cutoff). The bottom panel shows the gravitational wave likelihood for two fractional luminosity distance uncertainties. Gravitational waves can only be drawn from galaxies with true redshifts in the range $z_{\mathrm{lower}}\leq z\leq z_{\rm upper}$.}
    \label{fig:LGW_PGW}
\end{figure}

\color{Black}As mentioned in Section~\ref{sec:redMaGiC}, we apply a cut to the MICEcat catalogue to align with the redMaGiC sample, restricting the total possible redshift range to $z\in[0.15, 0.7]$. This prior is likewise applied to the redshift interpolant, $p_{\rm CBC}(z)$.\footnote{\color{black}Since our simulated events are drawn from this range, it serves as a natural prior for the analysis. In real data, however, restricting the redshift range could lead to artificially tight constraints by excluding events originating outside this range.} Consequently, gravitational wave events in this analysis are simulated with true redshifts between $z_{\rm lower}=0.15$ ($d_L\approx 720$~Mpc) and $z_{\rm upper}=0.7$ ($d_L\approx 4360$~Mpc). \color{Black}These events are then considered detectable when their observed luminosity distance falls below a threshold of $\hat{d}_L^{\mathrm{thr}} = 1550$~Mpc ($z\approx0.29$), which aligns with the current detection capabilities of the LIGO-Virgo-KAGRA (hereafter LVK) Collaboration \citep{GWTC3}. This choice of threshold is motivated by several reasons; firstly, while we anticipate a higher detection threshold in the future, we focus on the lower redshift regime as it is where we expect to get the most information on $H_0$. Secondly, the detection threshold must remain sufficiently low to ensure that no portion of the gravitational wave likelihood (as shown in Figure~\ref{fig:LGW_PGW}) extends past the realm where we have redshift data. To ensure this, we restrict our gravitational wave events to a fractional luminosity distance uncertainty of 10$\%$, a value based on the error from the dark siren GW190814 \citep{Abbott_2020} and representative of the projected error from the LVK network in upcoming years.

\color{Black}The probability of detecting a gravitational wave can be written
\begin{align}
    \label{eq:Heavy}
    p_{\mathrm{det}}^{\mathrm{GW}}(z,H_0) &=\int_{-\infty}^{\infty} \Theta(\hat{d}_L ; \hat{d}_L^{\mathrm{thr}}) \mathcal{L}_{\mathrm{GW}}(\hat{d}_L \mid d_L(z, H_0)) d \hat{d}_L\\
    \label{eq:pgw}
    &=\frac{1}{2}\left[1+\operatorname{erf}\left[\frac{\hat{d}_L^{\text {thr}}-d_L(z, H_0)}{\sqrt{2} \sigma_{d_L}}\right]\right],
\end{align}
\noindent where $\Theta\left(\hat{d}_L ; \hat{d}_L^{\mathrm{thr}}\right)$ is the Heaviside step function, which is unity when $\hat{d}_L \leq \hat{d}_L^{\mathrm{thr}}$ and 0 otherwise. The detection probability and gravitational wave likelihood functions can be seen plotted as a function of luminosity distance and redshift respectively in Figure~\ref{fig:LGW_PGW}. From the top panel of Figure~\ref{fig:LGW_PGW} as the fractional luminosity distance uncertainty approaches zero the detection probability reduces to a Heaviside step function (which can also be seen by looking at Equation~\ref{eq:Heavy}).
Finally, the probability that a merger, i.e. compact binary coalescence, CBC, will occur at redshift $z$ is given by 
\begin{equation}
    \label{eq:pcbcbig}
    p_{\mathrm{CBC}}(z)=\frac{p_{\text {rate}}(z) p_{\text {cat}}(z)}{\int_0^{\infty} p_{\text {rate}}(z) p_{\text {cat}}(z) d z}.
\end{equation}

Here, $p_{\text{cat}}$ represents the probability that a galaxy exists at a true redshift within the catalogue (however bear in mind we only have observed redshifts), and $p_{\text{rate}}$ is the probability of a galaxy at that redshift hosting a compact binary coalescence. \color{Black}Under this assumption, the $p_{\text{rate}}$ term drops out, and Equation~\ref{eq:pcbcbig} simplifies to
\begin{equation}
    p_{\mathrm{CBC}}(z)\approx p_{\text{cat}}(z \mid\{\hat{z}^i_g\}) =\frac{1}{N_{\text{gal}}} \sum_i^{N_{\text{gal}}} p_{\text{red}}(z \mid \hat{z}_g^i),
    \label{eq:pCBC}
\end{equation}
\noindent where $p_{\mathrm{red}}$ is the redshift posterior. If the galaxy redshifts are perfectly measured, Equation~\ref{eq:pCBC} reduces to a sum of Dirac delta distributions. However, we are interested in the effect of redshift uncertainties on the mean uncertainty in $H_0$. So, $p_{\text{red}}(z \mid \hat{z}_g^i)$ represents the fact that we are not able to measure the true redshifts of galaxies but instead we have observed redshifts:
\begin{equation}
    p_{\text{red}}(z \mid \hat{z}_g^i)=\frac{\mathcal{L}_{\text {red}}(\hat{z}_g^i \mid z) p_{\mathrm{bg}}(z)}{\int  dz \ \mathcal{L}_{\text{red}}(\hat{z}_g^i \mid z) p_{\mathrm{bg}}(z)}.
    \label{eq:zpost}
\end{equation}

Here, $p_\mathrm{{bg}}(z)$ is the prior on the redshift distribution and $\mathcal{L}_{\text {red}}$ is the likelihood used to generate observed redshifts from the true redshifts of each galaxy within the catalogue,
\begin{equation}
\label{eq:Lred}
    \mathcal{L}_{\text {red }}(\hat{z}_g^i \mid z)=\mathcal{N}(z\mid \sigma_z(1+z)+A).
\end{equation}

The standard deviation used in this likelihood takes the form of the relationship derived in Equation~\ref{eq:DESstdev}. A spectroscopic-like redshift is taken to have $\sigma_{z}=10^{-4}$ \citep{Carr}, while a photometric-like redshift is assigned $\sigma_{z}=10^{-2}$, which falls between the values reported by \citet{Cawthon_2022} and those found in Figure~\ref{fig:DESResidual}.
\begin{figure}[t!]
        \centering
    \includegraphics[width=1\linewidth]{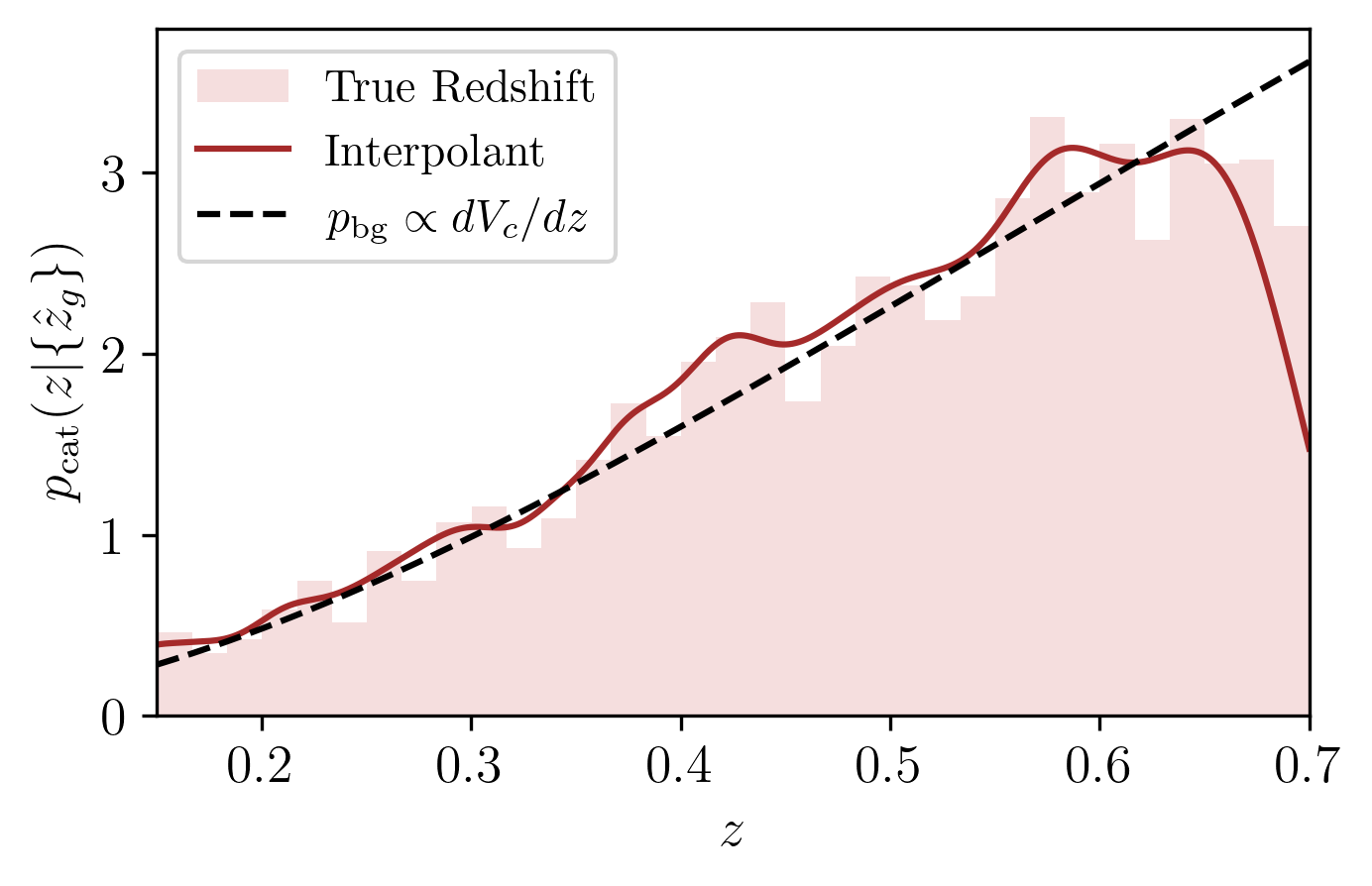}
    \caption{Probability of a true host galaxy redshift given an ensemble of measured redshifts for a 50~deg$^2$ localisation area and a photometric-like redshift uncertainty ($\sigma_{z}=10^{-2}$). The redshift bins show the distribution of true redshifts in the localisation area, and the dashed line shows the redshift prior (uniform-in-comoving-volume). The interpolant is given by $p_{\mathrm{CBC}(z)}$ (Equation~\ref{eq:pCBC}) for a single realisation.}
    \label{fig:RedshiftPriorInterp}
\end{figure}

In Equation~\ref{eq:zpost}, $p_{\mathrm{bg}}(z)$ is a prior on the redshift distribution that reflects our understanding of the background distribution of galaxies. The simplest choice we can make for this prior is to be uniform in comoving volume,
\begin{equation}
    p_{\mathrm{bg}}(z)=\frac{\frac{d V_c}{d z}}{\int_0^{\infty} \frac{d V_c}{d z} d z},
\end{equation}
\noindent where
\begin{equation}
    \frac{d V_c}{d z}=\frac{4 \pi}{E(z)}\left[\frac{c}{H_0}\right]^3\left[\int_0^z \frac{d z^{\prime}}{E\left(z^{\prime}\right)}\right]^2.
\end{equation}

\begin{figure}[t]
    \centering
    \includegraphics[width=0.95\linewidth]{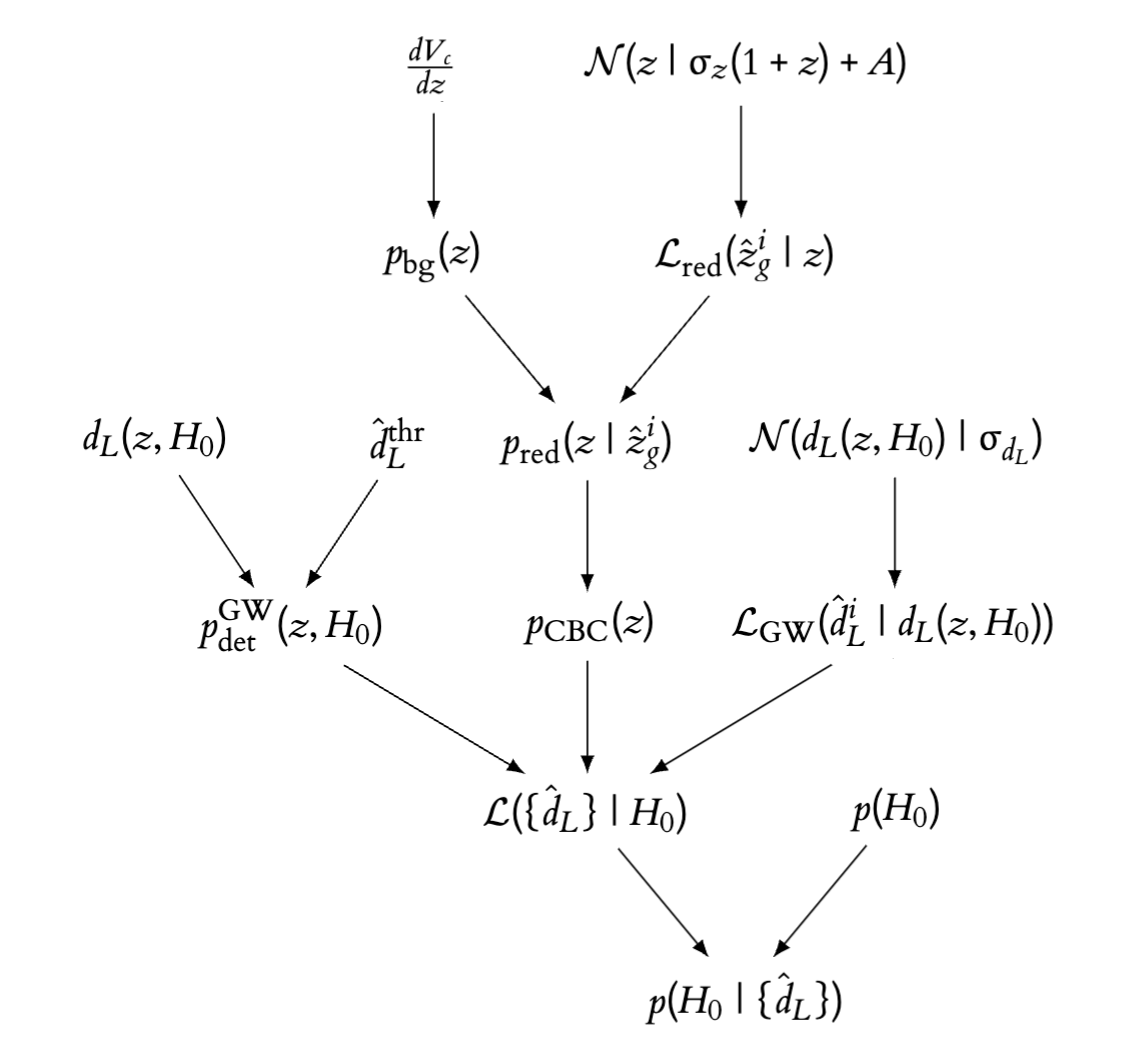}
    \caption{Schematic of the statistical framework used in this work.}
    \label{tikz:schematic}
\end{figure}

Figure~\ref{fig:RedshiftPriorInterp} shows the redshift posterior as described by Equation~\ref{eq:zpost}, alongside the uniform-in-comoving-volume prior and true redshift distribution. \color{Black}Figure~\ref{tikz:schematic} shows the schematic for the framework outlined in this section.

\subsection{Extensions to the model}
\label{sec:extensions}
In addition to the statistical framework described above and detailed in \citet{2023AJ....166...22G}, we introduce two key extensions: the ability to simulate different lines of sight within the MICEcat universe, and to filter galaxies within a specified localisation area. \color{Black}Localisation areas are constructed by randomly sampling right ascension and declination $(\alpha, \delta)$ positions from MICEcat, with each sample defining the central axis of a cone and thus the centre of a localisation region. Different lines of sight correspond to different choices of the central axis. \color{Black}These localisation areas are analogous to those provided by the LVK Collaboration, but are assumed to be circular in this analysis. We choose to represent three sky localisations throughout this analysis: 1, 10 and 50 square degrees. These values span the estimated median sky localisation for the ongoing fourth observing run (O4) \citep{Abbott_2020_prospects}, which is $41_{-6}^{+7}$~deg$^2$ for binary black hole mergers (90\% credible region). Additionally, 35--39\% of these events are expected to have a localisation region smaller than 20~deg$^2$. 

In general, the $H_0$ posterior for larger localisation areas tends to be less informative than smaller localisation areas, as the overdensities that provide crucial redshift information about the gravitational wave are more likely to be averaged out during the marginalisation over all galaxies. Indeed, when the localisation volume (which defines the three-dimensional region where potential host galaxies are distributed) exceeds a certain threshold, it approaches the homogeneity scale of the Universe. Beyond this scale, the large-scale structure no longer introduces statistical variations, and therefore localisation volumes that exceed the homogeneity scale will not contribute meaningful new information about $H_0$. A more detailed discussion with reference to dark siren analyses can be found in \ref{app:homo}.

\subsection{Nomenclature}
In the following sections we perform measurements across multiple realisations of multiple gravitational wave detections. To reduce ambiguity, we provide a summary of frequently used terms below.
\begin{enumerate}
    \item \textbf{Individual mean and standard deviation}: the mean and standard deviation of the $H_0$ posterior for any single realisation (line-of-sight), consisting of $N$ individual detected gravitational wave events.
    \item \label{ensmean}\textbf{Ensemble mean}: the average $H_0$ value calculated across all realisations. This quantity represents the overall mean measurement of $H_0$ we should expect. Note that each individual mean can differ from this `by chance', i.e. for observations in any given `universe' this is our expectation, but in reality our measurement may differ from the expectation due to noise.
    \item \textbf{Mean uncertainty}: the average standard deviation in $H_0$ for an individual realisation, averaged over all realisations. This quantity represents the typical uncertainty expected for a single realisation, but again, a given measurement of $H_0$ made from a single set of $N$ gravitational waves may have an uncertainty slightly better or worse due to random chance.
    \item \textbf{Standard error on the mean}: the standard deviation in the individual means, divided by the square root of the number of realisations. This value represents the variation in the individual mean $H_0$ values, reflecting the precision with which we can measure the ensemble mean in our analysis and identify any methodological biases from the truth.
    \item \textbf{Error on the error}: the standard deviation of uncertainties across all realisations. This quantifies the variability in our constraints on $H_0$ between different realisations, i.e. how the accuracy of our $H_0$ constraints could differ from `universe' to `universe' due to random chance.
\end{enumerate}
\color{Black}All subsequent results are based on 200 gravitational wave events, each along a unique line-of-sight direction, and 200 realisations.

\color{Black}\section{Results: Gaussian case}
\label{sec:resultsgaussian}

\begin{figure}[t!]
    \centering
    \includegraphics[width=1\linewidth]{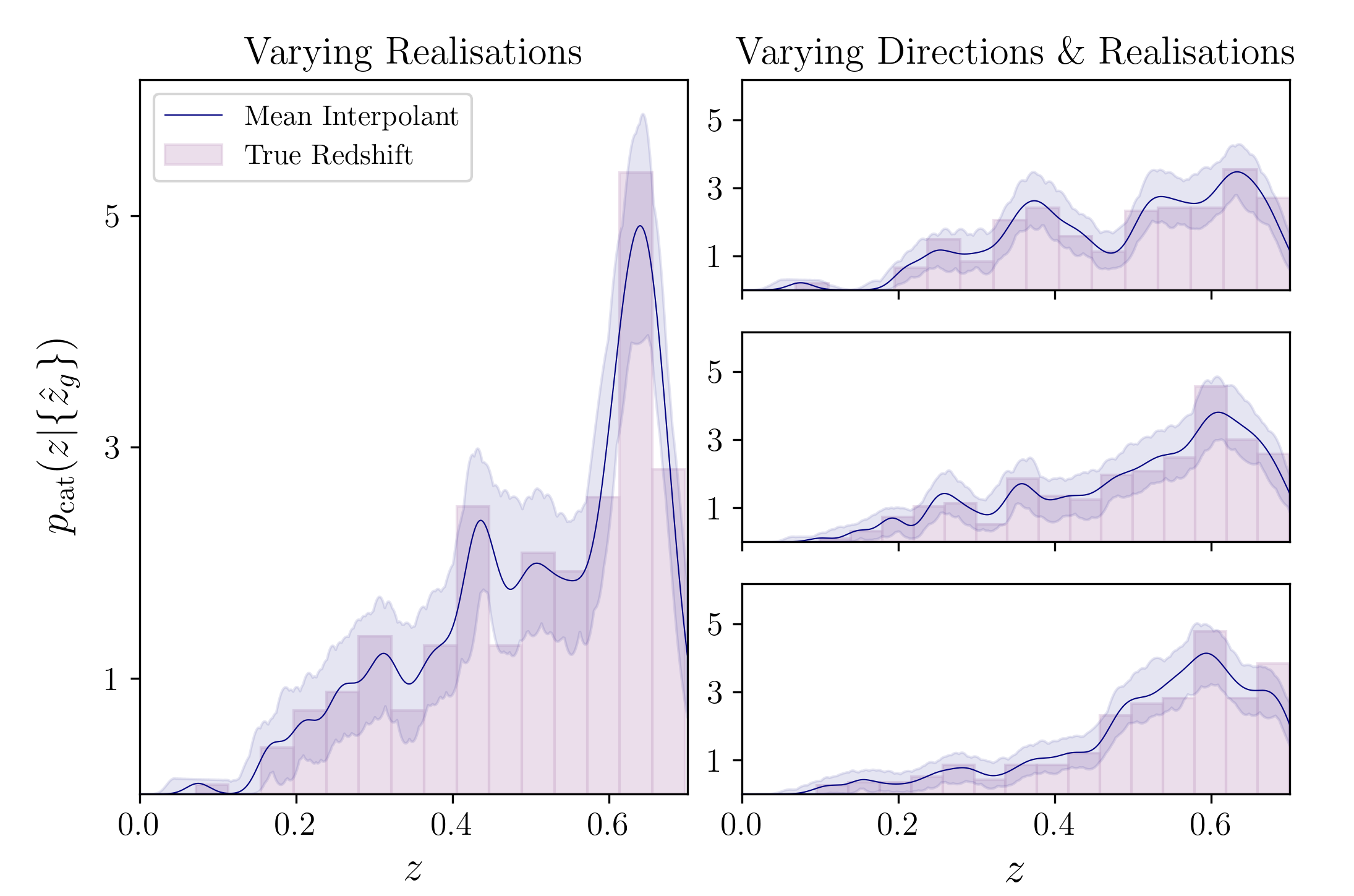}
    \caption{An illustration of how the redshift interpolant and the true redshift histogram vary across different realisations and directions. This interpolant corresponds to a localisation area of 1 deg$^2$ and $\sigma_z=10^{-2}$. The purple shaded region indicates the possible values the interpolant can take, while the pink histogram shows the true redshift distribution. The left panel shows how the interpolant varies for different realisations and in a single line of sight. The right panels show how the true redshift histogram and redshift interpolant vary for three different directions.}
    \label{fig:DirsRealsIllustration}
\end{figure}

\begin{table*}[t!]
    \centering
    \renewcommand{\arraystretch}{1.4}
    \resizebox{0.7\columnwidth}{!}{%
    \begin{tabular}{cccc}
        \textbf{} & $\bm{\sigma_z=10^{-4}}$ & $\bm{\sigma_z=10^{-3}}$ & $\bm{\sigma_z=10^{-2}}$ \\
        \midrule
        \textbf{1.0 deg$^2$ }  & $70.2 \pm 0.1$ \quad ($1.6 \pm 0.2$) & $70.1 \pm 0.1$\quad  ($1.6 \pm 0.3$) & $70.0 \pm 0.1$ ($1.9 \pm 0.2$) \\
        \textbf{10.0 deg$^2$ } & $70.1 \pm 0.2$ \quad ($2.3 \pm 0.3$) & $70.2 \pm 0.2$ \quad ($2.4 \pm 0.5$) & $70.2 \pm 0.2$ ($2.6 \pm 0.3$) \\
        \textbf{50.0 deg$^2$}  & $70.4 \pm 0.2$ \quad ($2.8 \pm 0.4$) & $70.6 \pm 0.2$ \quad ($2.9 \pm 0.5$) & $70.3 \pm 0.2$ ($2.9 \pm 0.4$) \\
        \bottomrule
        \end{tabular}}
    \caption{\color{Black}Mean and statistical uncertainties of $H_0$ for each combination of localisation area and redshift uncertainty. The values outside the brackets are the mean and standard error on the mean, and inside the brackets are the mean uncertainty and error on the error. All values have units of \hcon.}
    \label{tab:H0Sec5}
\end{table*}
We divide the results into three sections aligned with our three aims. As outlined in Section~\ref{sec:intro}, our first aim is to determine whether spectroscopic redshifts with Gaussian errors provide a better constraint on $H_0$ compared to photometric redshifts with Gaussian errors, and to identify the regime in which this effect is most pronounced. To achieve this, we generate observed redshifts from the true redshifts in MICEcat following Equation~\ref{eq:DESstdev}. In this simplified model, we set $A=0$, leaving the redshift uncertainty ($\sigma_z$) as the sole free parameter. We vary $\sigma_z$ from a spectroscopic- to photometric-like uncertainty ($10^{-4}$ to $10^{-2}$, respectively with an interim redshift uncertainty of $\sigma_z=10^{-3}$). For each combination of localisation area and redshift uncertainty, we simulate a single gravitational wave event in 200 directions, each yielding a posterior distribution for $H_0$. The combined effect across all 200 directions is found by taking the product of the posteriors from each direction (or equivalently, the sum of the log-posteriors). The result provides a constraint on $H_0$ that reflects what could realistically be achieved in the near future, as 200 gravitational wave events become an increasingly plausible target with upcoming LVK observation runs. We then run this entire experiment for 200 realisations (i.e. across 200 different `universes') to capture the variability in the posteriors as a result of Poisson noise. This approach differs from that of \citet{2023AJ....166...22G}, who generate 200 gravitational wave events along a single line-of-sight. This adjustment is motivated by practical considerations; any realistic sample will include gravitational waves from many lines-of-sight, and this has the advantage of avoiding any bias due to particular structures along one line-of-sight.

Each realisation is generated from the same sample of true redshifts (i.e. the same 200 lines-of-sight). Between each realisation, the following factors change:
\begin{enumerate}
    \item A new gravitational wave is generated (i.e. a new true luminosity distance is sampled).
    \item The observed parameters (specifically, $\hat{z}_g$ and $\hat{d}_L$) are resampled from the true parameters (redshift $z$ and luminosity distance $d_L$). 
\end{enumerate}

Varying directions involves sampling different lines-of-sight within the MICEcat universe. Since the large-scale structure changes, it is impossible to vary directions and keep the realisation constant. Thus, the following elements vary across different directions:

\begin{enumerate}
    \item All changes from varying realisations.
    \item A different distribution of true redshifts ($z$) due to the change in large-scale structure.
\end{enumerate}

Figure~\ref{fig:DirsRealsIllustration} shows how the redshift interpolant and true redshifts vary over different realisations and directions. By sampling different realisations we ensure that our results are independent of any particular noise realisation or gravitational wave event. By varying both realisations and directions, we can more accurately estimate the total uncertainty in $H_0$ by considering different patches of large-scale structure. This approach captures both the statistical uncertainty (from varying realisations) and the systematic uncertainty (due to sample variance) for any given noise realisation and direction.

\begin{figure*}
    \centering
    \includegraphics[width=0.9\linewidth]{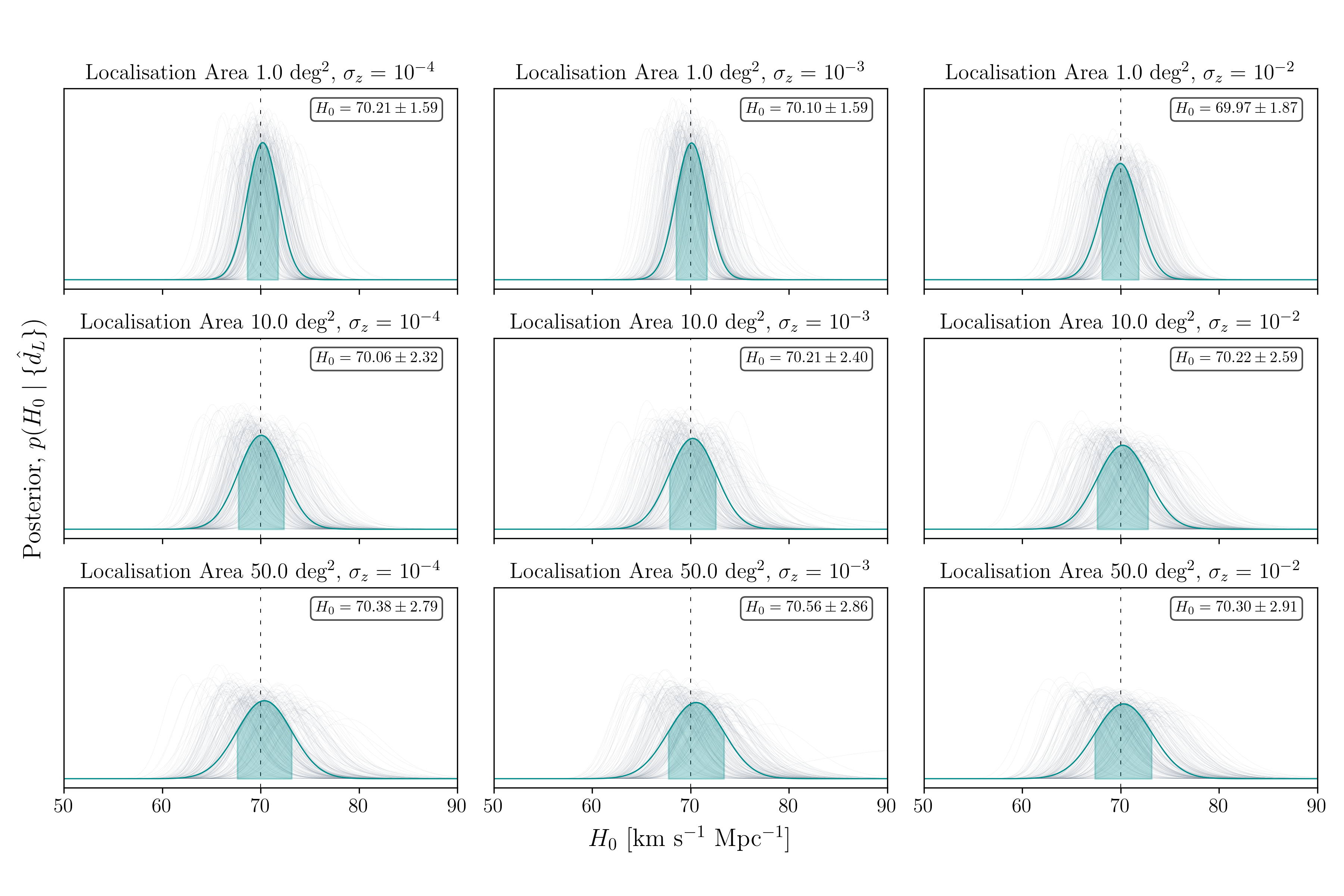}
    \caption{Each grey posterior represents a single noise realisation, generated from 200 distinct gravitational wave events each with their own line-of-sight direction. The black vertical line represents the true value of $H_0=70$ \hcon. The green Gaussian's show the ensemble mean and mean uncertainty across all grey posteriors. The ensemble mean and mean uncertainty are additionally annotated in the top right of each subplot (units are in \hcon).}
    \label{fig:AllPosteriors}
\end{figure*}
\begin{figure*}
    \centering
    \includegraphics[width=0.9\linewidth]{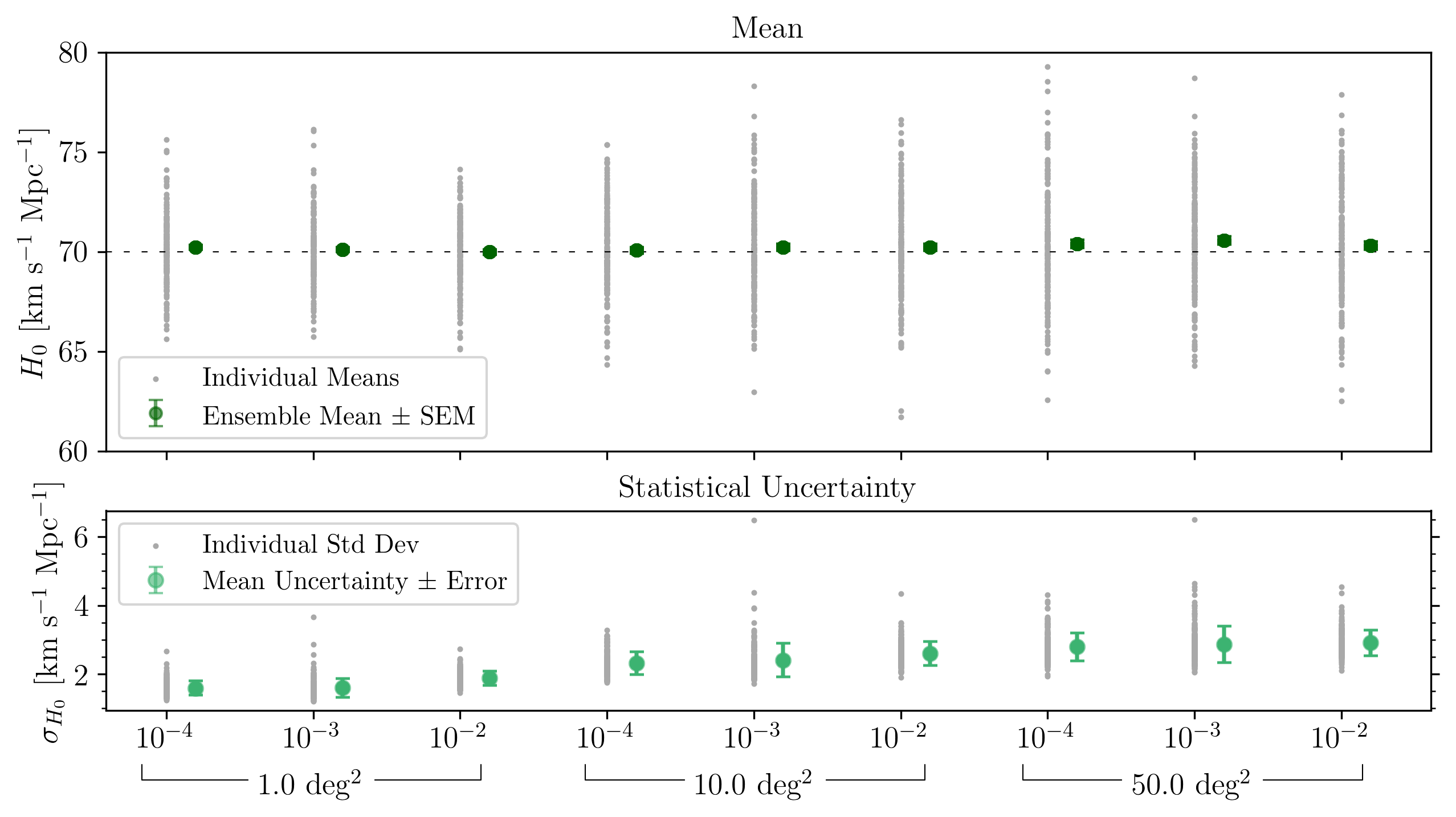}
    \caption{Predicted constraints on $H_{0}$ as a function of redshift error and localisation area. In the top panel, each grey point represents the individual mean $H_0$ values from each of the 200 realisations (corresponding to the mean of each grey posterior in Figure~\ref{fig:AllPosteriors}). The dark green points represent the ensemble mean and standard error on the mean. In the bottom panel, the grey points represent the individual standard deviations for each realisation. The light green points show the mean uncertainty, which reflects the average width of the posterior distributions, and the errorbars represent the error on the error.}
    \label{fig:VaryRealisDirs}
\end{figure*}

\color{Black}In Figure~\ref{fig:AllPosteriors} we show the resulting posteriors for each combination of localisation area and redshift uncertainty. The grey posteriors in each subplot show the the results from each of the 200 noise realisations, and the green Gaussian is generated with a mean and standard deviation given by the ensemble mean and mean uncertainty, respectively. In Figure~\ref{fig:VaryRealisDirs} we present the ensemble mean and standard error on the mean (top panel), along with the mean uncertainty and error on the error (bottom panel) across all 200 realisations. We do find evidence of a slight offset from the true value, $H_{0,\text{true}}=70$ \hcon, when considering the standard error on the mean. This offset, on the order of $\lesssim 0.3$ \hcon, is attributed to the partial overlap in the lines-of-sight within the 10 and 50 deg$^2$ localisation areas, leading to shared large-scale structure. Consequently, the error on the mean is likely an underestimation in these cases. Given the small magnitude of this offset relative to the typical uncertainty from a single realisation of 200 gravitational wave events, we do not consider this bias significant. \color{Black}Although sampling non-overlapping cones would provide a straightforward way to mitigate this offset, the limited sky coverage of MICEcat prevents this for the two largest localisation areas, where covariance and thus error underestimation is most significant. \color{Black}Future work could instead address this by repeating the experiment with multiple independent simulation realisations or taking into account the covariance between overlapping lines-of-sight. 

Table~\ref{tab:H0Sec5} provides a summary of the results from this analysis. We find that reducing the localisation area from 50 to 1 deg$^2$ yields a relative reduction in the mean uncertainty by a factor of $\sim 1.8$ for spectroscopic-like redshifts and $\sim 1.5$ for photometric-like redshifts. Spectroscopic-like redshifts consistently yield smaller mean uncertainties than photometric-like redshifts, with the greatest improvement --- a factor of 1.2 reduction in uncertainty --- observed in the 1 deg$^2$ localisation area. This highlights that spectroscopic redshifts are more impactful for smaller localisation areas, as there is a negligible reduction in mean uncertainty between photometric and spectroscopic redshifts for the 50 deg$^2$ localisation area.

\subsection{Additional Tests}
The results in Figure~\ref{fig:VaryRealisDirs} show that the mean uncertainty in $H_0$ decreases with reductions in either the localisation area or the redshift uncertainty. However, these changes are relatively modest. To understand the regime in which we would expect a reduction in either localisation area or redshift uncertainty to have more a pronounced effect on the mean uncertainty in $H_0$, we conduct additional stress-tests.

\subsubsection{When do smaller localisation areas provide the most benefit?}
\label{sec:extensiontest1}
The first test aims to determine the conditions under which a smaller localisation area has the greatest impact on the mean uncertainty in the Hubble constant, $\sigma_{H_0}$. Specifically, we examine two scenarios: varying the number of gravitational wave events (each observed along a different line of sight) at a fixed fractional luminosity distance uncertainty of 10\%, and varying the fractional luminosity distance uncertainty while maintaining a fixed number of 200 events. Both scenarios are conducted using the methodology described earlier in this section. \color{Black}For the latter case, we note that although gravitational waves with larger distance uncertainties often also have poorer sky localisation, the purpose of this analysis is to isolate the source of any improvements. For this reason, we vary the luminosity distance uncertainty and the localisation area independently.

\color{Black}We find the former scenario always results in a consistent $\sim 1.8\times$ improvement when going from 50 to 1 deg$^2$, for event counts ranging from 50 to 200. In contrast, the second scenario --- reducing the fractional luminosity distance uncertainty --- demonstrates a much more pronounced enhancement in constraining power when the localisation area is decreased from 50 to 1 deg$^2$. The results of this test are presented in Table~\ref{tab:dl_loc}.

Table~\ref{tab:dl_loc} demonstrates a clear increase in the relative improvement of mean $H_0$ uncertainties for localisation areas of 1 and 50 deg$^2$ as the fractional luminosity distance uncertainty decreases. This suggests that the benefits of enhanced sky localisation diminish as radial uncertainty, rather than angular uncertainty (from the localisation area), becomes the primary source of error. While smaller localisation areas reduce the number of potential host galaxies, large fractional luminosity distance uncertainties exacerbate the degeneracy between redshift and luminosity distance, leading to broader constraints on $H_0$.

\begin{table}[t]
    \centering
    \renewcommand{\arraystretch}{1.2}
    \resizebox{\columnwidth}{!}{%
    \begin{tabular}{cccc}
    \textbf{}& \textbf{$\bm{\sigma_{d_L}=10\%}$} & \textbf{$\bm{\sigma_{d_L}=5\%}$} & \textbf{$\bm{\sigma_{d_L}=1\%}$}  \\
    \midrule
    \textbf{50 deg$^2$} & $\pm 2.79$ & $\pm1.52$ &  $\pm0.34$\\
    \textbf{1 deg$^2$} & $\pm1.59$ & $\pm0.60$ & $\pm0.07$ \\
    \hline
    \textbf{Relative Improvement} & $1.75\times$ & $2.53\times$ & $4.86\times$  \\
    \bottomrule
    \end{tabular}}
    \caption{Mean uncertainty in $H_0$ (in units of \hcon) for three values of fractional luminosity distance uncertainty, evaluated for localisation areas of 1 and 50 deg$^2$ (constant $\sigma_z=10^{-4}$). The bottom row indicates the relative improvement between the uncertainties from the two localisation areas.}
    \label{tab:dl_loc}
\end{table}

\subsubsection{When do spectroscopic redshifts provide the most benefit?}
\label{sec:extensiontest2}
The second test aims to determine the conditions under which spectroscopic redshifts provide the most benefit.  
Similar to Section~\ref{sec:extensiontest1}, we reduce the fractional luminosity distance uncertainty and compare the differences in $H_0$ uncertainties between spectroscopic- and photometric-like redshift catalogues for each of the three $\sigma_{d_L}$ values.

The results in Table~\ref{tab:dl_specz} lead to conclusions similar to those in Section~\ref{sec:extensiontest1}, emphasising that high fractional luminosity distance uncertainty significantly restricts the advantages of using spectroscopic redshifts. The issue is not that spectroscopic redshifts lack utility; rather, the high fractional luminosity distance uncertainty limits the extent to which their benefits can be realised. 

\color{Black}The uncertainty in luminosity distance primarily arises from its strong correlation with the binary system’s orbital inclination. This degeneracy can be mitigated through the detection of higher-order modes in the gravitational wave signal, enabling a more precise inference of the inclination and, consequently, a reduction in the uncertainty of the luminosity distance \citep{2021ApJ...912L..10C}. \color{Black}Upcoming third-generation detectors like the Einstein Telescope (ET; \citealt{2010CQGra..27s4002P}) or the Cosmic Explorer (CE; \citealt{reitze2019cosmicexploreruscontribution}) will offer significantly improved sensitivity, enabling the detection of these higher-order harmonics. For events below $z=0.1$ ($\approx 462$ Mpc), incorporating these modes is expected to improve sky localisation by 5–25\% and distance measurements by 20–45\% \citep{Borhanian_2020}. Leveraging the advanced capabilities of these future detectors, spectroscopic redshifts are anticipated to impose much tighter constraints on $H_0$ compared to photometric redshifts.
\begin{table}[t]
    \centering
    \renewcommand{\arraystretch}{1.2}
    \resizebox{\columnwidth}{!}{%
    \begin{tabular}{cccc}
    \textbf{} & $\bm{\sigma_{d_L}=10\%}$ & $\bm{\sigma_{d_L}=5\%}$ & $\bm{\sigma_{d_L}=1\%}$ \\
    \midrule
   \textbf{photo-$z$} & $\pm1.86$ & $\pm1.01$ & $\pm0.58$ \\
    \textbf{spec-$z$} & $\pm 1.59$ & $\pm0.60$ &  $\pm0.07$\\
    \hline
    \textbf{Relative Improvement} & $1.16\times$ & $1.68\times$ & $8.29\times$  \\
    \bottomrule
    \end{tabular}}
    
    \caption{The mean uncertainty in $H_0$ (in units of \hcon) for three fractional luminosity distance uncertainties, considering both photometric and spectroscopic redshift uncertainties (fixed localisation area of 1 deg$^2$). The bottom row shows the relative improvement between spectroscopic- and photometric-like redshift catalogues for each $\sigma_{d_L}$.}
    \label{tab:dl_specz}
\end{table}

This section focused on addressing our first aim; to determine whether the increased precision of spectroscopic redshifts leads to tighter constraints on $H_0$ and whether these improvements depend on the localisation area. We confirm that the high precision spectroscopic redshifts ($\sigma_z=10^{-4}$) consistently reduce the mean uncertainty in $H_0$ compared to lower precision photometric redshifts ($\sigma_z=10^{-2}$). Additionally, this improvement was found to depend on the size of the localisation area, with spectroscopic redshifts providing the greatest benefit in smaller regions. For a 1 deg$^2$ localisation area, spectroscopic-like redshifts were found to reduce the mean uncertainty by a factor of 1.2, whereas for a 50 deg$^2$ localisation area, the reduction is negligible. Moreover, this reduction is also highly sensitive to the fractional luminosity distance uncertainty, with larger uncertainties reducing the benefits of spectroscopic redshifts. Based on these findings, spectroscopic follow-up of current gravitational wave events should be prioritised for those with the smallest fractional luminosity distance uncertainties, as this factor has the greatest influence on the benefits of higher-precision redshifts.

\color{black}The discrepancy between our results and those of \citet{Borghi_2024} --- who report a degradation in $H_0$ precision by a factor of $\sim 3$ when using photometric redshifts, compared to our factor of $\sim 1.2$ --- can be attributed to several methodological differences. The most significant difference is that their assumed photometric redshift uncertainties are five times larger than those used here, naturally leading to a reduction in precision and a corresponding larger difference between spectroscopic and photometric $H_0$ constraints.\footnote{\color{black}While they do additionally use a more conservative spectroscopic redshift uncertainty of $\sigma_z=10^{-3}$, compared to our $\sigma_{z}=10^{-4}$, results from Section~5 show that the difference between these two is minimal --- at this level of precision, the dominant contribution to the $H_0$ uncertainty comes from the luminosity distance precision rather than redshift uncertainty.} We have verified using our pipeline that increasing our photometric redshift to match that of \citet{Borghi_2024} raises the relative difference from approximately 1.2 to 2.1. The remaining discrepancy is likely because \citet{Borghi_2024} jointly perform a galaxy catalogue and spectral siren analysis to obtain their results. Doing so will likely impact constraints on $H_0$ using a photometric catalogue more severely than a spectroscopic one, as redshift uncertainties propagate not only through the redshift–distance relation but also through the transformation from source-frame to detector-frame masses.\footnote{\color{black}The source-frame and detector-frame masses are related by $m_{\rm det}=m_{\rm source}(1+z)$.}.

\color{Black}\section{Results: realistic case}
In addressing our first aim, we assumed Gaussian errors for both spectroscopic and photometric redshifts, thereby excluding the impact of outlier redshifts --- cases where the observed and true redshift significantly differ. In reality, photometric redshift catalogues are highly prone to outliers due to the inherent limitations of the methods used to collect them. Photometric redshifts are estimated using a limited set of broadband filters, which can introduce ambiguities when different redshifts produce similar observed colours \citep{2001A&A...368...74M}. This degeneracy can lead to significant errors in redshift estimation. Additionally, the accuracy of photometric redshifts relies heavily on the spectral energy density (SED) templates used; if these templates fail to accurately represent the observed galaxies, the resulting redshift estimates can be incorrect. This motivates our second aim, which is to explore how outliers in photometric catalogues may bias the measurement of $H_0$. To address this, we adopt a methodology similar to that outlined in Section~\ref{sec:resultsgaussian}, but now compare two methods for generating observed redshifts. Previously, we generated observed redshifts from the true MICEcat redshifts using Equation~\ref{eq:DESstdev} with $A=0$ and a varying redshift uncertainty, $\sigma_z$. For this section, our two observed redshifts are generated as follows:
\begin{enumerate}
    \item Observed redshifts are generated from true redshifts using Equation~\ref{eq:Lred} with $\sigma_z=0.019$ and $A=-0.013$, which models the general relationship between redMaGiC photometric and spectroscopic redshifts (as shown in Figure~\ref{fig:DESResidual}). As with Section~\ref{sec:resultsgaussian}, there will be no outliers in this sample. We call this method `Gaussian $z_{\mathrm{obs}}$'.
    \item Observed redshifts are generated by crossmatching the MICEcat and redMaGiC redshifts, selecting the redMaGiC galaxies whose spectroscopic redshifts are closest to the MICEcat redshifts. The observed redshifts are taken to be the corresponding redMaGiC photometric redshifts (and photometric redshift uncertainties). We refer to this method as `redMaGiC $z_{\mathrm{obs}}$'. \color{Black}This method will introduce a number of outliers consistent with a realistic photometric redshift catalogue. For redMaGiC, the outlier fraction ---  defined as $|z_{\rm spec}-z_{\rm photo}|>3\sigma_z$ --- is approximately constant across the three localisation areas, with a mean value of $\sim6\%$.
\end{enumerate}

\label{sec:resultsrealistic}
\begin{figure}[t!]
    \centering
    \includegraphics[width=\linewidth]{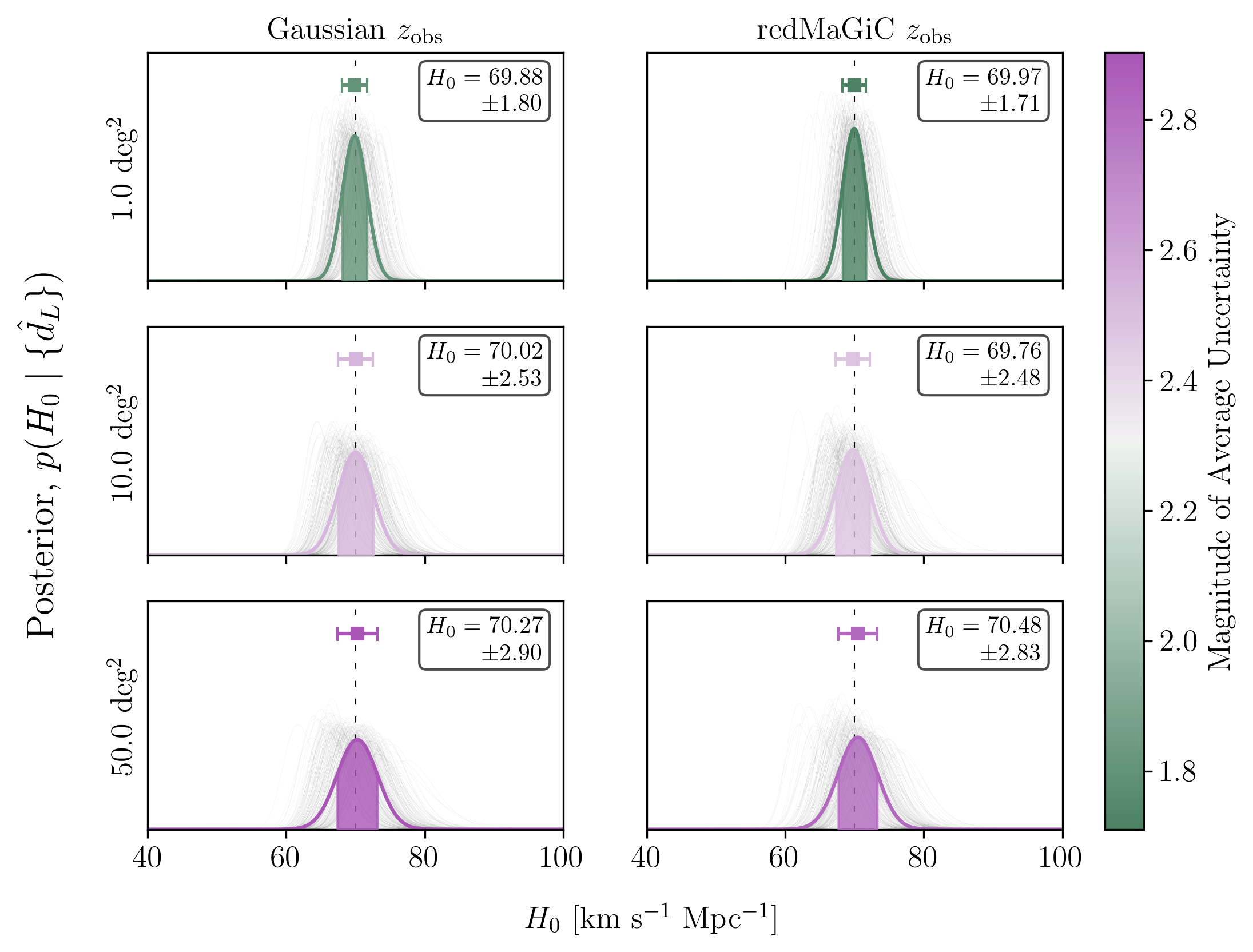}
    \caption{Combined posteriors for a single gravitational wave event over 200 directions, with 200 noise realisations. In the left column, the observed redshifts are generated from Equation~\ref{eq:Lred} with standard deviation given by Equation~\ref{eq:DESstdev} with $A=-0.013$ and $\sigma_z=0.019$. In the right column, the redMaGiC photometric redshifts are used as the observed redshifts. Each coloured Gaussian represents the ensemble mean and mean uncertainty calculated over all 200 realisations, with the colour reflecting the magnitude of the mean uncertainty. Scatter points are overlaid to visually reinforce the ensemble mean and mean uncertainty, which are also annotated in the top-right corner of each subplot for clarity (units in \hcon).}
    \label{fig:ComparisonPostDES}
\end{figure}
Since we only use the redMaGiC catalogue to derive observed redshifts from the sample of spectroscopic redshifts, there is no need to account for either the incompleteness of the redMaGiC sample, or the selection bias associated with the Luminous Red Galaxy sample. Given that our Gaussian $z_{\mathrm{obs}}$ method is shown to be a good fit to the errors in Figure~\ref{fig:DESResidual}, the only difference between redMaGiC and Gaussian observed redshifts will be higher order redshift dependencies or non-Gaussianities (outliers) in the redMaGiC data.

With these two sets of observed redshifts, we generate posteriors as in Section~\ref{sec:resultsgaussian}; a single gravitational wave event is generated in 200 directions, over 200 noise realisations. The resulting posteriors for the two sets of observed redshifts and three localisation areas is shown in Figure~\ref{fig:ComparisonPostDES}.

Despite the outliers present in the redMaGiC sample, we consistently retrieve an unbiased measurement of $H_0$ for each localisation area, except for the slight bias in the standard error on the mean caused by overlapping cones in the 50 deg$^2$ localisation area. While we find the mean uncertainty across all three localisation areas to be slightly lower for the redMaGiC observed redshifts than for the Gaussian observed redshifts, the error on the error shown in Table~\ref{tab:Stats} points towards this discrepancy being statistically insignificant. Furthermore, consistent with the findings in Section~\ref{sec:resultsgaussian}, the mean uncertainty for both types of observed redshifts increases with the size of the localisation area.

For our second aim, we conclude that photometric redshift catalogues that contain redshift outliers do not bias the measurement of $H_0$. \color{black}While the inclusion of outliers leads to a slight reduction in the mean uncertainty, the difference in $H_0$ uncertainty between redMaGiC and MICEcat is not statistically significant; as shown in Table~\ref{tab:H0Sec5}, the difference remains below $1\sigma$ across the 200 realisations. \color{Black}Consequently, photometric catalogues can serve as a reliable foundation for conducting dark siren analyses, with no risk of exacerbating the Hubble tension.

\begin{table}[t!]
\centering
\renewcommand{\arraystretch}{1.2}
\resizebox{\columnwidth}{!}{%
\begin{tabular}{lcc}
\small
& {\textbf{Gaussian $z_{\mathrm{obs}}$}} 
& {\textbf{redMaGiC $z_{\mathrm{obs}}$}} \\
\midrule
\textbf{1.0 deg$^2$ } & $69.9 \pm 0.1$ ($1.8 \pm 0.2$) & $70.0 \pm 0.1$ ($1.7 \pm 0.2$) \\
\textbf{10.0 deg$^2$} & $70.0 \pm 0.2$ ($2.5 \pm 0.4$) & $69.8 \pm 0.2$ ($2.5 \pm 0.5$) \\
\textbf{50.0 deg$^2$} & $70.3 \pm 0.2$ ($2.9 \pm 0.4$) & $70.5 \pm 0.2$ ($2.8 \pm 0.5$) \\
\bottomrule
\end{tabular}}
\caption{Table of means and statistical uncertainties for the Gaussian and redMaGiC observed redshifts. The `mean' column represents the ensemble mean and the standard error on the mean, whereas the `statistical uncertainty' column shows the mean uncertainty and error on the error. The slight bias in the standard error on the mean for the 50 deg$^2$ localisation area is attributed to overlapping large-scale structure, causing the error to be underestimated (as discussed in Section~\ref{sec:resultsgaussian}).}
\label{tab:Stats}
\end{table}
\section{Results: uniform sub-sampling}
\label{sec:completeness}
Our final aim is to determine the completeness level at which the advantage of using spectroscopic redshifts is outweighed by the loss of accuracy due to an incomplete galaxy sample. To achieve this, we generate observed redshifts with a methodology that aligns with Section~\ref{sec:resultsgaussian}. Given that spectroscopic redshift catalogues are more prone to high levels of incompleteness, we use only a spectroscopic-like redshift uncertainty ($\sigma_z=10^{-4}$) in this scenario. The redshift interpolant (Equation~\ref{eq:pcbcbig}) currently only accounts for an in-catalogue component and thus assumes that the host galaxy of the gravitational wave event is always present in the galaxy catalogue. \color{Black}To address incompleteness, we adjust the redshift interpolant to also account for an out-of-catalogue component, following a simplified approach to the method outlined in \citet{Chen:2017rfc, Finke:2021aom}. \color{Black}The complete redshift interpolant for both in- and out-of-catalogue parts is:
\begin{align}
    \label{eq:compl}
    p_{\mathrm{CBC}}(z)&=f_{\mathrm{comp}}p_{\mathrm{in}}+(1-f_{\mathrm{comp}})p_{\mathrm{out}}\\
    &=\frac{f_\mathrm{comp}}{N_{\text {gal}}}\left( \sum_i^{N_{\text{gal}}} p_{\text{red}}(z \mid \hat{z}_g^i)\right)+(1-f_{\mathrm{comp}}) p_{\mathrm{bg}}(z),
\end{align}
\noindent where $f_{\mathrm{comp}}$ is the completeness fraction of the catalogue. When the catalogue is 100\% complete, Equation~\ref{eq:compl} reduces to the original interpolant, as only the in-catalogue galaxies contribute to the probability distribution. For incomplete catalogues ($f_{\mathrm{comp}}<1$), the information from the in-catalogue galaxies is increasingly dampened by the contribution from the out-of-catalogue background distribution from the redshift prior, $p_{bg}(z)$. 

\begin{figure}[t!]
    \centering
    \includegraphics[width=1\linewidth]{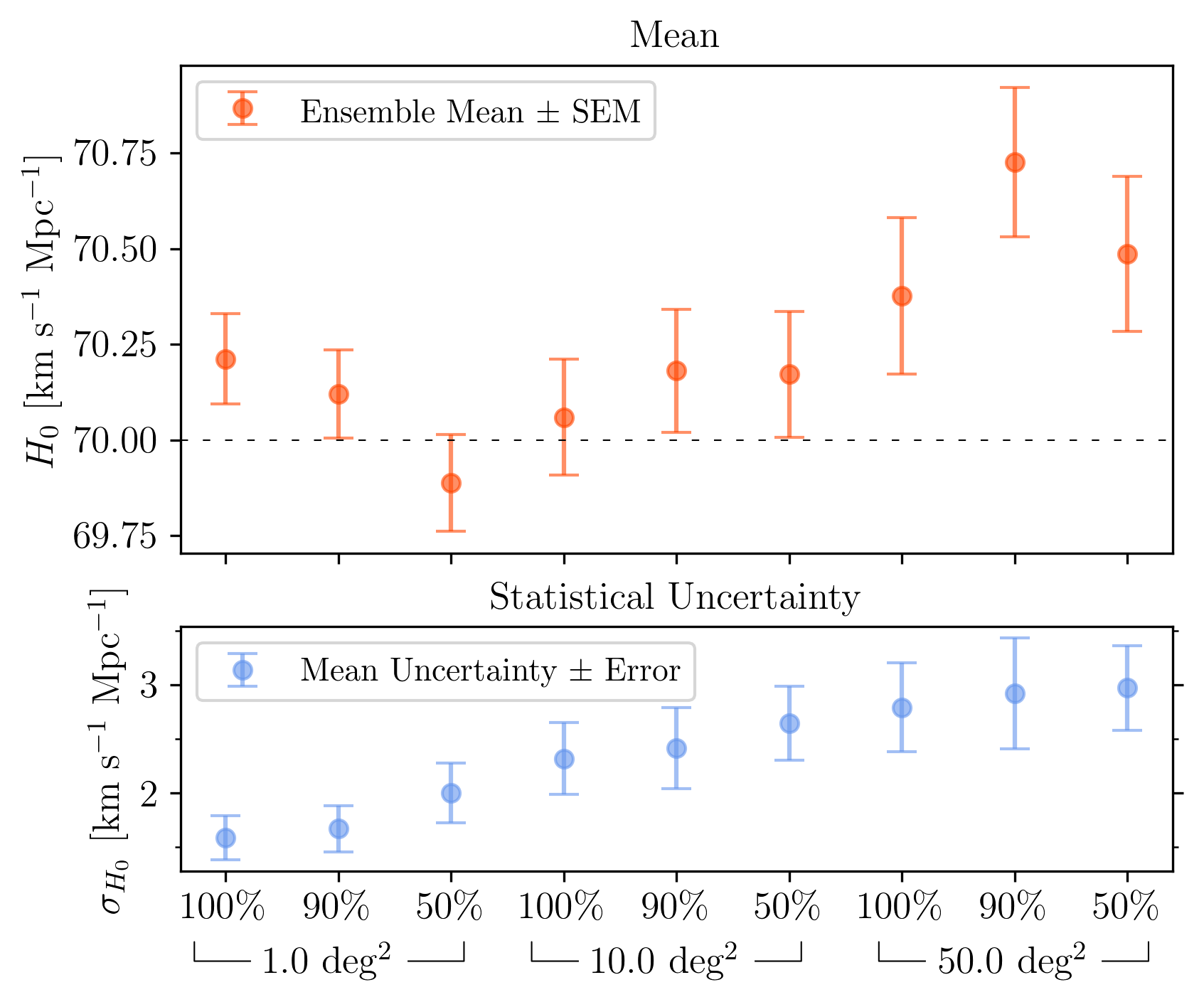}
    \caption{Predicted constraints on $H_0$ as a function of localisation area and catalogue completeness fraction (constant spectroscopic-like redshift uncertainty of $\sigma_z=10^{-4}$). The top panel shows the mean and standard error on the mean for each combination of localisation area and completeness fraction. The bottom panel shows the mean uncertainty and error on the error.}
    \label{fig:completeness}
\end{figure}

To simulate an incomplete catalogue, we generate a gravitational wave using the full redshift catalogue, and remove a percentage of the galaxies that corresponds to the completeness fraction. While in reality incomplete catalogues will not be uniformly sub-sampled, but instead follow a selection function, the methodology required to take into account such selection functions is substantially more complex (see \citealt{gray2023jointcosmologicalgravitationalwavepopulation}), and so we do not consider this in this work. Since we expect that in removing a fraction of the galaxies the shape of the redshift interpolant (as shown in Figure~\ref{fig:RedshiftPriorInterp}) will remain mostly unchanged, the dominant effect should be an increase in the mean uncertainty as the completeness fraction decreases, as information in the form of galaxies is being removed. Figure~\ref{fig:completeness} shows the results for 3 localisation areas and 3 completeness fractions: 100\% (which are the same results as in Section~\ref{sec:resultsgaussian}), 90\% and 50\%. As aforementioned, we use a spectroscopic redshift uncertainty of $\sigma_z=10^{-4}$ and generate observed redshifts with the same methodology as Section~\ref{sec:resultsgaussian} (with 200 directions, each with a single event and over 200 noise realisations). 
The slight bias present in the standard error on the mean (errorbars in the top panel of Figure~\ref{fig:completeness}) can once again be attributed to the correlation arising from overlapping cones. Overall, we find that a lower completeness fraction leads to a higher mean uncertainty, with this effect being more pronounced for smaller localisation areas. For a 1 deg$^2$ localisation area, reducing the completeness from 100\% to 50\% increases the mean uncertainty by a factor of 1.3. In contrast, the same reduction in completeness for a 50 deg$^2$ localisation area results in only a 1.1-fold increase, highlighting the greater sensitivity of smaller localisation regions to catalogue completeness.

A spectroscopic catalogue with a completeness fraction of 50\% seems to yield a slightly larger mean uncertainty than having a fully complete photometric catalogue without outliers (using the results from Section~\ref{sec:resultsgaussian} and Figure~\ref{fig:VaryRealisDirs}). Additionally, for small localisation areas ($<10$ deg$^2$), having a 90\% complete catalogue of spectroscopic redshifts yields similar results to having a fully complete photometric catalogue (including outliers). This highlights that while spectroscopic redshifts can improve the constraints on $H_0$, their effectiveness heavily depends on maintaining a high completeness fraction, which, as noted in Section~\ref{sec:motivation}, is resource-intensive to achieve. Therefore, the most effective approach to obtaining an unbiased and precise measurement of $H_0$ will involve using a combination of both spectroscopic and photometric redshifts, at least until more complete spectroscopic catalogues become available. This approach, however, will introduce additional complexity in the selection function.

\section{Conclusions}
\label{sec:conclusions}
We have explored the dark siren method for measuring the Hubble constant, focusing on the impact of precision redshifts on the uncertainty associated with a measurement of $H_0$. For our first aim, we used Gaussian errors to generate observed redshifts from a catalogue of true redshifts (given by MICEcat), simulating both spectroscopic- and photometric-like uncertainties. We obtained unbiased measurements of $H_0$ in each case, and found that spectroscopic redshifts consistently lead to smaller uncertainties in $H_0$, with the effect being most pronounced in smaller localisation areas. 

For our second aim, we examined the influence of redshift outliers on the analysis (outliers being defined as cases where there is a significant discrepancy between the true and observed redshift). We performed a similar analysis to our first aim, but instead compared two methods of generating observed redshifts. The first generates observed redshifts from the true redshifts using Gaussian errors that follow the same distribution as the redMaGiC sample. The second uses the redMaGiC redshifts as the observed redshifts. The former method will not include outliers, but the latter will; given that both methods will generate redshifts that follow the same average trend, any differences we see in our results can be attributed to the presence of these outliers. We found that the redMaGiC observed redshifts (with outliers) did not bias the measurement of $H_0$ in any case. While the mean uncertainty was found to be marginally smaller for the redMaGiC observed redshifts, this was not found to be statistically significant given the standard deviation of these measurements. Overall, while using photometric catalogues with realistic outlier fractions to measure $H_0$ is shown to provide an unbiased measurement, the precision associated with spectroscopic redshifts will still provide a tighter constraint on $H_0$.

For our final aim, we explored the effects of uniformly sub-sampling the redshift catalogue after generating the gravitational wave event. While this approach aimed to simulate an incomplete spectroscopic redshift catalogue, it does not reflect the reality that incompleteness typically follows a selection function, which was beyond the scope of this analysis. Nevertheless, we show that uniform sub-sampling does not introduce any bias into the measurement of $H_0$. As expected, a decrease in the completeness fraction resulted in an increase in the mean uncertainty in $H_0$. This effect was found to be most prominent in smaller localisation areas, where there are fewer galaxies. While Section~\ref{sec:resultsgaussian} found that spectroscopic-like redshifts were most valuable in smaller localisation areas, Section~\ref{sec:completeness} found these smaller localisation areas are the most affected by incompleteness, highlighting a trade-off between the benefits of higher precision and the challenges posed by incomplete catalogues.

Overall our results suggest that while having spectroscopic redshifts will always result in a more precise estimate of $H_0$, using photometric redshifts will not bias the $H_0$ measurement. However, it is important to note that we rely exclusively on very high-quality photometric redshifts, so bias and larger redshift uncertainties may be present in analyses that use less ideal catalogues. While we find that uniform sub-sampling primarily increases mean uncertainty in a measurement of $H_0$, incorporating more realistic cases of incompleteness through a selection function (as could be done with \texttt{gwcosmo} from \citealt{gray2023jointcosmologicalgravitationalwavepopulation}) would provide clearer insights into the effects of spectroscopic redshift catalogue incompleteness on these measurements; doing this would further clarify the role of spectroscopic redshifts in dark siren analyses.

In light of these findings and the exposure times detailed in Section~\ref{sec:motivation}, we conclude that, under the conditions examined in this work and with current gravitational wave detection capabilities, spectroscopic redshifts offer only minimal gains in constraining $H_0$. Our results suggest that when photometric redshifts are available, spectroscopic follow-up is only justified for events with the smallest fractional luminosity distance errors and somewhat justified for particularly well localised events. In all other scenarios, the added precision from spectroscopic data does not outweigh the associated costs. However, the magnitude of the fractional luminosity distance uncertainty still remains a key factor limiting the benefits of spectroscopic redshifts. With 3G detectors on the horizon, their full potential is only beginning to be realised. At that point, spectroscopic redshifts will emerge as an indispensable tool for achieving precise measurements of the Hubble constant using dark sirens, potentially offering critical insights into the origins of the Hubble tension and its implications for cosmology.

\begin{acknowledgement}
We would like to thank Chris Lidman and Chris Blake for their contributions and insightful questions that enriched this analysis. We would also like to thank Liana Rauf for her valuable feedback on earlier versions of this manuscript.
\end{acknowledgement}

\paragraph{Funding Statement}
This research was conducted by the Australian Research Council Centre of Excellence for Gravitational Wave Discovery (project number CE230100016) and funded by the Australian Government.

%\paragraph{Competing Interests}

\paragraph{Data Availability Statement}
This analysis is adapted from \citet{2023AJ....166...22G}. The MICEcat lightcone simulation \citep{Fosalba_2015} can be found from \url{https://cosmohub.pic.es/home}. The redMaGiC Dark Energy Survey sample \citep{Rozo_2016,2005astro.ph.10346T} can be found from \url{https://www.darkenergysurvey.org/the-des-project/data-access/}. Code to reproduce plots is available on request.

%\endnote in some journals will behave like \footnote; and \printendnotes will not output anything. 
\printendnotes

\appendix
\section{Homogeneity scale}
\label{app:homo}
To determine the localisation volume of a gravitational wave without any luminosity distance information, consider a spherical sector with an opening angle, $\theta$, and radius, $r$.\footnote{A spherical sector can be described as the union of a spherical cap and the cone formed by the centre of the sphere and the base of the cap.} The localisation volume in this case is given by:
\begin{align}
    V_{\text{spherical sector}}=\frac{2}{3}\pi r^3(1-\cos\theta).
\end{align}

\begin{figure}[t!]
    \centering    
    \includegraphics[width=\linewidth]{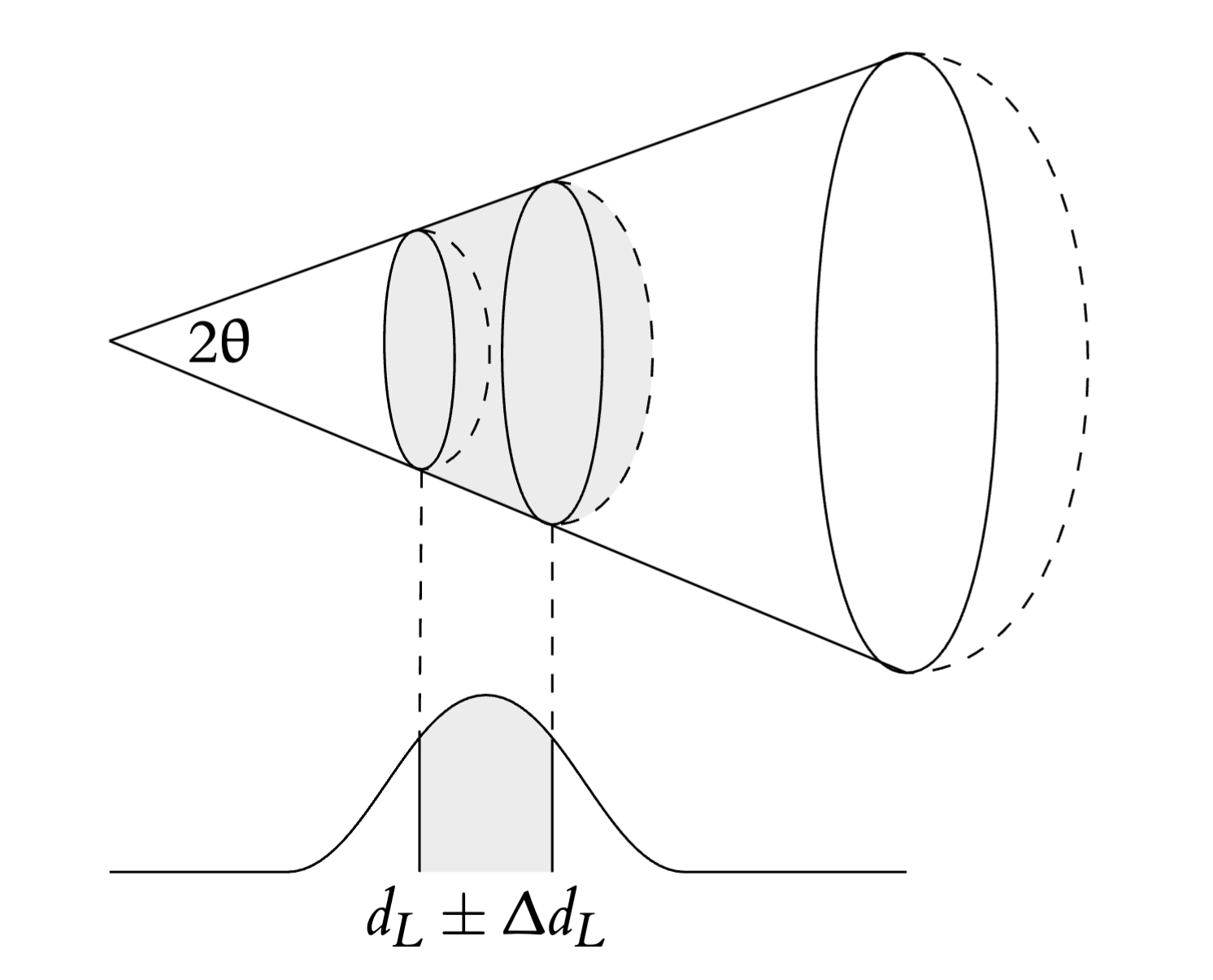}
    \caption{Diagram illustrating the comoving volume shell created by the luminosity distance posterior and opening angle (which is related to the localisation area).}
    \label{fig:cone}
\end{figure}

In performing a dark siren analysis, we already make use of the distance posterior, which provides a central luminosity distance and error. This information can be used to further constrain the localisation volume. Making use of the two luminosity distance bounds ($d_L\pm\Delta d_L$) and converting them to comoving distances, we can define a comoving volume shell (see Figure~\ref{fig:cone})
\begin{align}
V_{\text{shell}}=
\frac{2}{3}\,\pi 
\left(r_{\text{upper}}^{3}- r_{\text{lower}}^{3}\right)
\left(1 - \cos\theta\right),
\label{eq:comovingvolumeshell}
\end{align}
\noindent where $r_{\text{upper}}$ and $r_{\text{lower}}$ are the comoving distances corresponding to the upper and lower luminosity distance bounds, respectively. The opening angle, $\theta$, is related to the localisation area by converting it to a solid angle and computing:
\begin{align}
\theta = \arccos\left( 1 - \frac{\Omega_{\text{sr}}}{2\pi} \right).
\end{align}

The comoving volume shell described by Equation~\ref{eq:comovingvolumeshell} can be visualised in Figure~\ref{fig:cone} as the shaded region. 
With this information, we can approximate the localisation volume for a given localisation area and gravitational wave luminosity distance posterior, and compare it to the volume of a sphere with a radius of the homogeneity scale. 

The homogeneity scale ($\mathcal{R}_H$) marks the transition where the distribution of galaxies in the Universe becomes statistically uniform, beyond which the localisation volume is expected to contribute no new information about $H_0$. In such cases, the uncertainty in $H_0$ should plateau to the standard deviation of the prior distribution (which in our case is the standard deviation of a flat distribution between 40 and 100 \hcon). Current literature (notably, \citealt{Scrimgeour_2012, Ntelis_2017}) measure $\mathcal{R}_H$ to be $\sim 100$~Mpc h$^{-1}$, we quote this value with an uncertainty of $\pm 20$~Mpc h$^{-1}$. 

\begin{figure}[t!]
    \centering
    \includegraphics[width=1\linewidth]{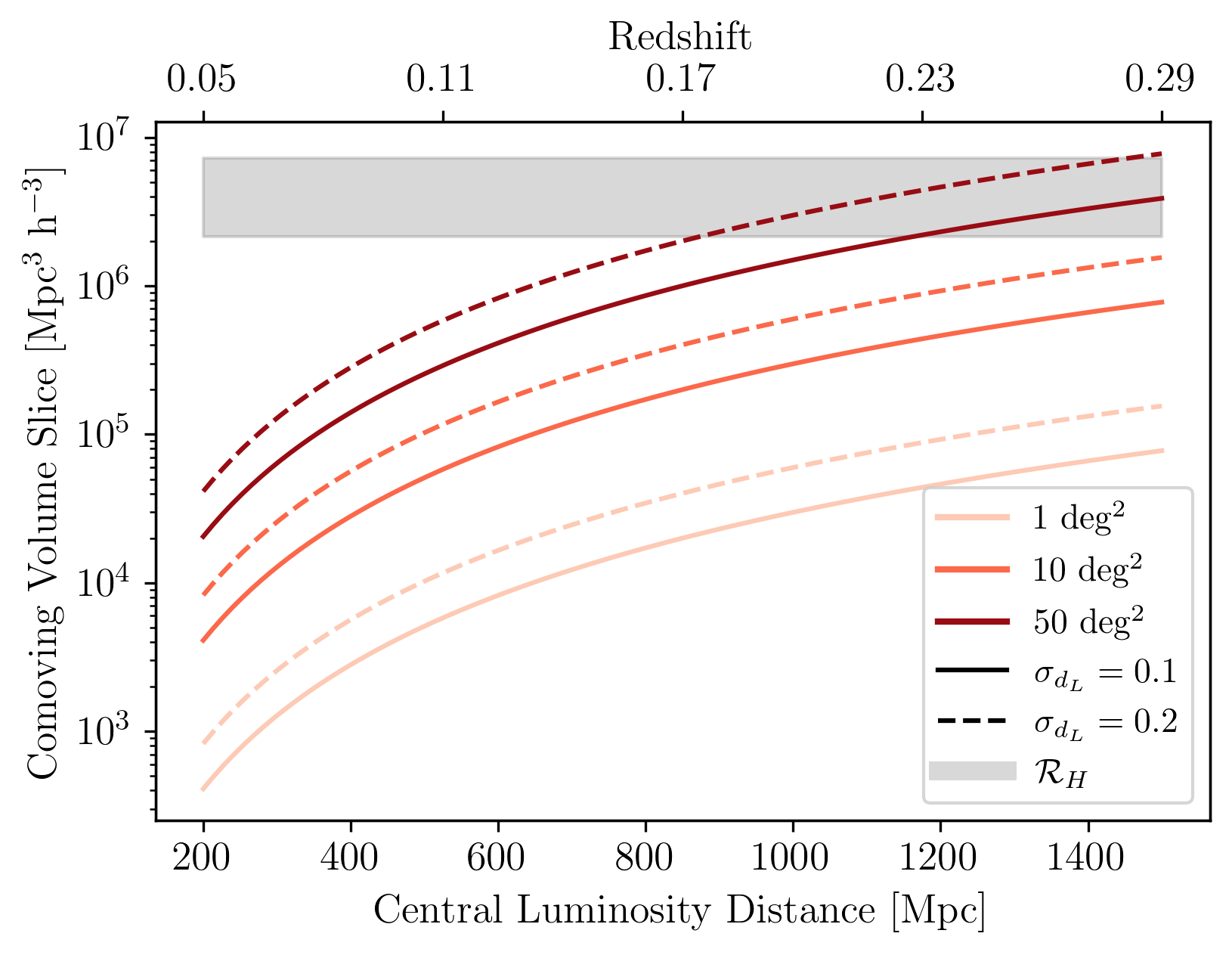}
    \caption{Volume of a comoving shell as a function of central luminosity distance. The central luminosity distance is the mean of the luminosity distance posterior shown in Figure~\ref{fig:cone}. The three coloured lines represent three values of the localisation area, and the two line styles show the value of the fractional luminosity distance uncertainty (which will widen the posterior in Figure~\ref{fig:cone}). The grey shaded region indicates the volume of a sphere with radius given by the homogeneity scale.}
    \label{fig:homoscale}
\end{figure}

We vary both the localisation area --- by adjusting the opening angle, $\theta$ --- and the luminosity distance (both $d_L$ and $\Delta d_L$) to determine when the resulting localisation volume reaches the homogeneity scale. Figure~\ref{fig:homoscale} presents the results for the three localisation areas used in the main work, and two fractional luminosity distances, $\sigma_{d_L}=0.1$ and 0.2. 

We find that for localisation areas of 1 and 10 deg$^2$, the corresponding localisation volumes (for both values of $\sigma_{d_L}$) remain below the homogeneity scale for all gravitational wave luminosity distances considered in this work. This indicates that these events provide meaningful contributions to the measurement of $H_0$.

In contrast, for a 50 deg$^2$ localisation area, the localisation volume exceeds the homogeneity scale at high luminosity distances in both $\sigma_{d_L}$ cases. As a result, the information gain on $H_0$ becomes negligible for gravitational waves at these distances, making inclusion of such events in $H_0$ estimates unwarranted.

To test this hypothesis --- that $H_0$ measurements cease to be informative as the localisation volume surpasses the homogeneity scale --- we modify the selection function (Equation~\ref{eq:pgw}) to allow gravitational waves to be drawn from specific luminosity distance bins, $[d_{L,\min}, d_{L,\max}]$:
\begin{align}
    p_{\mathrm{det}}^{\mathrm{GW}}(z,H_0)
&=\frac{1}{2}\left[\operatorname{erf}\left[\frac{\hat{d}_L^{\text {max}}-d_L(z, H_0)}{\sqrt{2} \sigma_{d_L}}\right]\right.\nonumber\\
&\qquad\left.-\operatorname{erf}\left[\frac{\hat{d}_L^{\text {min}}-d_L(z, H_0)}{\sqrt{2} \sigma_{d_L}}\right]\right].
\end{align}

We generate 200 events distributed across five equally spaced luminosity distance bins between 720 and 1550~Mpc. The lower bound, 720 Mpc, corresponds to $z\sim 0.15$ --- our minimum redshift --- while 1550~Mpc represents the original threshold luminosity distance.

\begin{figure}[t!]
    \centering
    \includegraphics[width=1\linewidth]{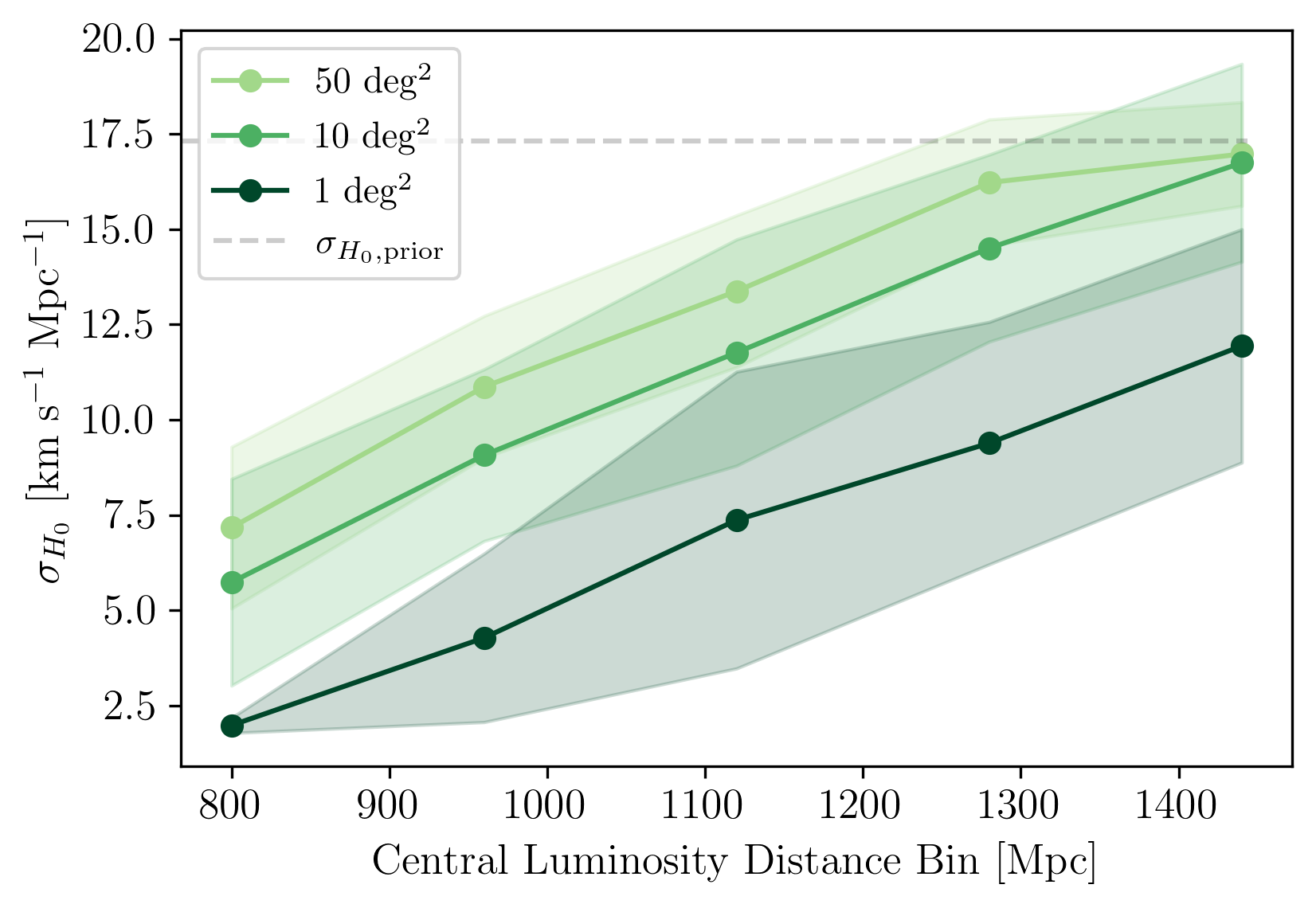}
    \caption{Uncertainty in a measurement of $H_0$ as a function of the luminosity distance of the gravitational waves, shown for three localisation areas (constant $\sigma_z=10^{-4}$, $\sigma_{d_L}=0.1$). Each posterior is generated from 200 events, each in a different direction. The shaded regions represent the mean and standard deviation over 20 realisations. The dashed line indicates the standard deviation of the prior distribution, which is $\mathcal{U}$(40,100) \hcon in this analysis.}
    \label{fig:dlbin}
\end{figure}

Each of the 200 events is, as before, generated from an individual line-of-sight, and the combined posterior is the product of the individual posteriors across all events. We conduct this experiment for the original three localisation areas (with a constant $\sigma_z=10^{-4}$ and $\sigma_{d_L}=0.1$) and run 20 realisations (i.e. repeat the experiment 20 times for each luminosity distance bin and localisation area). The results, averaged over the realisations, are shown in Figure~\ref{fig:dlbin}. 

We observe a clear trend of increasing $H_0$ uncertainty with gravitational wave distance, driven by the expanding localisation volume and the corresponding increase in the number of potential host galaxies. As expected, the two smaller localisation areas (1 and 10~deg$^2$) show a steady increase in $H_0$ uncertainty as the gravitational wave luminosity distance bin increases. However, for the largest localisation area (50~deg$^2)$, the uncertainty appears to plateau around $\sim1300~$Mpc, aligning with the scale at which the 50 deg$^2$ localisation area with $\sigma_{d_L}=0.1$ reaches the homogeneity scale in Figure~\ref{fig:homoscale}. Additionally, the value at which the $H_0$ uncertainty plateaus corresponds to the standard deviation expected for a flat posterior within the range defined by the $H_0$ prior.\footnote{The standard deviation expected for a flat posterior between two prior values is $\sigma_{H_0} = \sqrt{\langle x^2 \rangle - \langle x \rangle^2}$, 
where $\langle x^2 \rangle=\int_{H_0^{\text{min}}}^{H_0^{\text{max}}}\frac{x^2 \,dx}{H_0^{\text{max}} - H_0^{\text{min}}}$ and $\langle x \rangle = \int_{H_0^{\text{min}}}^{H_0^{\text{max}}} \frac{x \,dx}{H_0^{\text{max}} - H_0^{\text{min}}}$. This evaluates to $17.3$ \hcon when the prior width is $60$ \hcon.} This indicates that the measurement is no longer contributing meaningful information beyond what was already encoded in the prior.  At this point, the Universe is statistically homogeneous, and the structure that initially contributes both information and uncertainty to the measurement of $H_0$ ceases to be informative.

These findings not only serve as a strong proof of concept but also underscore that when the localisation volume exceeds the homogeneity scale, the statistical information from large-scale structure is effectively lost, providing no further improvement to constraints on the Hubble constant.

\bibliography{example}
\balance

%\printbibliography
\end{document}